\documentclass[draft]{agujournal2019}
\usepackage{url} 
\usepackage{lineno}
\usepackage{soul}
\usepackage{amsmath} 
\usepackage{rotating}

\newcommand{\dg}{$^{\circ}$}
\newcommand{\kms}{km~s$^{-1}$}

\draftfalse

\journalname{Space Weather}

\begin{document}

%
%

\newcommand{\actaa}{Acta Astron.} 
\newcommand{\araa}{Annu. Rev. Astron. Astrophys.} 
\newcommand{\aar}{Astron. Astrophys. Rev.} 
\newcommand{\ab}{Astrobiol.} 
\newcommand{\aj}{Astron. J.} 
\newcommand{\apj}{Astrophys. J.} 
\newcommand{\apjl}{Astrophys. J. Lett.} 
\newcommand{\apjs}{Astrophys. J. Suppl. Ser.} 
\newcommand{\ao}{Appl. Opt.} 
\newcommand{\apss}{Astrophys. Space Sci.} 
\newcommand{\aap}{Astron. Astrophys.} 
\newcommand{\aapr}{Astron. Astrophys. Rev.} 
\newcommand{\aaps}{Astron. Astrophys. Suppl.} 
\newcommand{\baas}{Bull. Am. Astron. Soc.} 
\newcommand{\caa}{Chinese Astron. Astrophys.} 
\newcommand{\cjaa}{Chinese J. Astron. Astrophys.} 
\newcommand{\cqg}{Class. Quantum Gravity} 
\newcommand{\gal}{Galaxies} 
\newcommand{\gca}{Geochim. Cosmochim. Acta} 
\newcommand{\icarus}{Icarus} 
\newcommand{\jcap}{J. Cosmol. Astropart. Phys.} 
\newcommand{\jgr}{J. Geophys. Res.} 
\newcommand{\jgrp}{J. Geophys. Res.: Planets} 
\newcommand{\jqsrt}{J. Quant. Spectrosc. Radiat. Transf.} 
\newcommand{\memsai}{Mem. Soc. Astron. Italiana} 
\newcommand{\mnras}{Mon. Not. R. Astron. Soc.} 
\newcommand{\nat}{Nature} 
\newcommand{\nastro}{Nat. Astron.} 
\newcommand{\ncomms}{Nat. Commun.} 
\newcommand{\nphys}{Nat. Phys.} 
\newcommand{\na}{New Astron.} 
\newcommand{\nar}{New Astron. Rev.} 
\newcommand{\physrep}{Phys. Rep.} 
\newcommand{\pra}{Phys. Rev. A} 
\newcommand{\prb}{Phys. Rev. B} 
\newcommand{\prc}{Phys. Rev. C} 
\newcommand{\prd}{Phys. Rev. D} 
\newcommand{\pre}{Phys. Rev. E} 
\newcommand{\prl}{Phys. Rev. Lett.} 
\newcommand{\psj}{Planet. Sci. J.} 
\newcommand{\planss}{Planet. Space Sci.} 
\newcommand{\pnas}{Proc. Natl Acad. Sci. USA} 
\newcommand{\procspie}{Proc. SPIE} 
\newcommand{\pasa}{Publ. Astron. Soc. Aust.} 
\newcommand{\pasj}{Publ. Astron. Soc. Jpn} 
\newcommand{\pasp}{Publ. Astron. Soc. Pac.} 
\newcommand{\rmxaa}{Rev. Mexicana Astron. Astrofis.} 
\newcommand{\sci}{Science} 
\newcommand{\sciadv}{Sci. Adv.} 
\newcommand{\solphys}{Sol. Phys.} 
\newcommand{\sovast}{Soviet Ast.} 
\newcommand{\ssr}{Space Sci. Rev.} 
\newcommand{\uni}{Universe} 

%
%

\title{ARCANE - Early Detection of Interplanetary Coronal Mass Ejections}

%
%

\authors{Hannah T. Rüdisser\affil{1,2}, Gautier Nguyen\affil{3}, Justin Le Louëdec\affil{1}, Emma E. Davies\affil{1}, Christian Möstl\affil{1}}

\affiliation{1}{Austrian Space Weather Office, GeoSphere Austria, Graz, Austria}
\affiliation{2}{Institute of Physics, University of Graz, Graz, Austria}
\affiliation{3}{DPHY, ONERA, Université de Toulouse, 31000, Toulouse, France}

\correspondingauthor{H. T. Rüdisser}{hannah@ruedisser.at}


\begin{keypoints}
\item We provide a modular framework to develop and evaluate methods for early detection of interplanetary coronal mass ejections in real-time.

\item We assemble an archive of real-time solar wind data to assess models under realistic operational conditions.

\item We reliably detect high-impact events in a real-time setting and achieve acceptable performance on low-impact events.
\end{keypoints}

%
%

\begin{abstract}
Interplanetary coronal mass ejections (ICMEs) are major drivers of space weather disturbances, posing risks to both technological infrastructure and human activities. Automatic detection of ICMEs in solar wind in situ data is essential for early warning systems. While several methods have been proposed to identify these structures in time series data, robust real-time detection remains a significant challenge. In this work, we present ARCANE - the first framework explicitly designed for early ICME detection in streaming solar wind data under realistic operational constraints, enabling event identification without requiring observation of the full structure. Our approach evaluates the strengths and limitations of detection models by comparing a machine learning-based method to a threshold-based baseline. The ResUNet++ model, previously validated on science data, significantly outperforms the baseline, particularly in detecting high-impact events, while retaining solid performance on lower-impact cases. Notably, we find that using real-time solar wind data instead of high-resolution science data leads to only minimal performance degradation. Despite the challenges of operational settings, our detection pipeline achieves an F1-Score of $0.37$, with an average detection delay of 24.1\% of the event´s duration while processing only a minimal portion of the event data. As more data becomes available, the performance increases significantly. These results mark a substantial step forward in automated space weather monitoring and lay the groundwork for enhanced real-time forecasting capabilities.
\end{abstract}

\section*{Plain Language Summary}

Solar storms are major drivers of the weather in space, which can disrupt technology and impact our daily lives on Earth. Early warning systems rely on the ability to automatically recognize these events in data from satellites near Earth, but current methods still have significant limitations. In this study, we present a new framework to evaluate how well different methods can automatically detect solar storms while observing the solar wind. We compare a machine learning model to a simpler threshold-based method. Our results show that the machine learning model performs much better, especially for identifying the most impactful events. Additionally, it still handles less critical events effectively. This system is a step forward for automated space weather monitoring and helps to improve real-time forecasting and early warning capabilities.

%
%

\section{Introduction}

Interplanetary coronal mass ejections (ICMEs) are among the primary drivers of space weather disturbances. These massive eruptions from the Sun occur more frequently during solar maximum \cite{richardson_2012} and are responsible for the most intense geomagnetic storms \cite{echer_2013}. Since their discovery in the $1970$s \cite{gosling_1973, burlaga_1981, klein_1982}, ICMEs have been the subject of extensive research, with numerous studies focusing on their properties, propagation, and geoeffectiveness \cite<e.g.>[]{kilpua_2017}. These efforts have greatly advanced our understanding of their physical properties and have led to a number of publications of event catalogs, providing start and end times of ICMEs, as observed by various spacecraft at L1 \cite{jian_2006, lepping_2006, richardson_2010, Chi_2016, moestl_2017, nguyen_automatic_2019, moestl2020_icmerate, nguyen_multiclass_2025}.

ICMEs are typically characterized by an enhanced, smoothly rotating magnetic field, a declining solar wind velocity, and low plasma beta ($\beta$), where $\beta$ is the ratio of thermal to magnetic pressure. They are often preceded by shocks and turbulent sheath regions, while their main structure is generally considered to be a magnetic cloud or flux rope. However, detecting ICMEs remains challenging due to the variability of their in situ signatures and the complex solar wind environment \cite<e.g.>[]{zurbuchen_2006, Chi_2016, kilpua_2017, al-haddad_2025_cme_repr}. The observed signatures can vary significantly depending on factors such as the spacecraft trajectory through the ICME, interactions with other CMEs, or the presence of additional transient solar wind structures, such as stream interaction regions (SIRs) \cite<e.g.>{kilpua2009, good2018correlation, Lugaz_2018, salman2020radial, davies_2022_multispacecraft, ruedisser_2024}.

Manually identifying ICME signatures is time-consuming and prone to inconsistencies. Catalogs often differ significantly, with studies showing that only a subset of the ICMEs in one catalog are present in another \cite<e.g.>[]{richardson_2014_catdifferences, rudisser_automatic_2022}. This inconsistency is a major challenge for the development of automatic detection methods, as they rely on expert-labeled data for training and validation. At the same time, automatically detecting these events in solar wind in situ data is essential for early warning systems, needed to mitigate the impact of space weather on critical infrastructure. A reliable ICME detection method could serve as a real-time trigger for more computationally intensive analysis, or even be deployed onboard a spacecraft to start different observational tasks. 

Several approaches have been proposed to automate ICME detection. Traditional methods, such as threshold-based techniques \cite{lepping_2005_automaticdetection}, algorithms based on the Grad-Shafranov reconstruction technique \cite{Hu_2018}, and spatio-temporal entropy analysis \cite{ojedagonzalez2017}, have been employed. Regardless, these approaches are highly dataset-dependent and often struggle to generalize across the diverse in situ signatures of ICMEs. More recent advancements in machine learning offer a promising alternative.

Early machine learning approaches for solar wind classification and ICME detection employed methods such as Gaussian Process classification and simple Convolutional Neural Networks, demonstrating the feasibility of automated in situ data analysis \cite{camporeale_classification_2017, nguyen_automatic_2019, li_categorization_2020}. Subsequent advancements have leveraged deep learning architectures, such as UNets, to reframe ICME detection as a time series segmentation task \cite{rudisser_automatic_2022, chen_ru-net_2022}. More recent efforts have explored alternative machine learning techniques, including feature selection with random forests to classify magnetic flux ropes \cite{farooki_machine_2024}, probabilistic neural networks for identifying solar wind structures \cite{narock_classifying_2024}, and supervised classification pipelines to further refine the automatic detection of magnetic flux ropes and solar transients \cite{pal_automatic_2024}.

The latest advancements integrate object detection frameworks inspired by the YOLO family \cite{redmon_2016_yolo}, enabling efficient multi-class event detection with minimal post-processing. These approaches unify the identification of ICMEs and SIRs within a single model, achieving high Precision and Recall in extensive data sets \cite{nguyen_multiclass_2025}.

Despite the substantial advancements in automatic detection of large-scale structures in solar wind in situ data, significant challenges remain in the field. One of the primary complications is the inherent subjectivity and inconsistency in existing event catalogs. As highlighted in previous work, the identification of ICMEs is still heavily dependent on expert visual labeling, which is time-consuming and highly biased \cite<e.g.>[]{rudisser_automatic_2022}.

In addition to the challenges stemming from incomplete catalogs, there are several difficulties when applying automatic detection methods in operational real-time settings. One potential issue is the difference in data quality. While many of the models discussed above are trained, validated, and tested on science-quality data, real-time data is often of lower quality, with increased noise and potentially containing more gaps. Another challenge arises from the fact that many current detection methods rely on analyzing the entire time series before identifying an ICME and setting its boundaries. Yet, in an operational setting, the goal is to detect events as early as possible with sufficient confidence to avoid false alarms. This means that the model must be capable of identifying initial signs of an ICME, such as the shock, sheath, or the early onset of the magnetic obstacle (MO), even when only part of the data is available.

In this study, we introduce ARCANE (Automatic Real-time deteCtion ANd forEcast), a comprehensive and modular machine learning framework designed for the real-time detection, prediction, and analysis of ICMEs using solar wind in situ data. By integrating multiple state-of-the-art methods across different stages, ARCANE provides an automated, data-driven and physics-informed system for space weather forecasting and operational early warning.

Here, we focus on the first component of ARCANE: the early detection of ICMEs. In contrast to \citeA{rudisser_automatic_2022}, as well as other related studies \cite<e.g.>[]{camporeale_classification_2017, nguyen_automatic_2019, pal_automatic_2024}, which evaluated ICME detection retrospectively, we assess our model under realistic operational conditions, emphasizing evaluation methodologies specifically designed to measure its capability to detect ICMEs in a real-time scenario.

The structure of this paper is as follows: Section~\ref{sec:data} describes the data sets used in this study, including in situ solar wind measurements, event catalogs, and the generation of labels. Section~\ref{sec:framework} outlines the detection module of ARCANE, detailing model architecture, training process, and the baseline model we compare to. Additionally, we introduce postprocessing and evaluation strategy and validate this approach. Section~\ref{sec:results} presents our experimental results, comparing the performance of our model against the baseline and assessing its early detection capabilities. Finally, in Section~\ref{sec:conclusion}, we discuss the implications of our findings for operational space weather forecasting and outline possible directions for future research.

\section{Data}\label{sec:data}

\subsection{In Situ Data}\label{sec:insitu}

In previous work \cite{rudisser_automatic_2022}, we utilized solar wind in situ data from Wind, STEREO-A and STEREO-B, with varied input parameters. \citeA{li_categorization_2020}, \citeA{nguyen_multiclass_2025} and many other machine learning studies made use of the OMNI data set \cite{King_Papitashvili_2005}, as it provides preprocessed, 1-min-resolution data, specifically intended to support studies of the effects of solar wind variations on the magnetosphere and ionosphere. Data is available from 1995 onward. Nonetheless, its lack of real-time availability renders it unsuitable for operational space weather forecasting.

\begin{figure}
\begin{center}
\includegraphics[width=\textwidth]{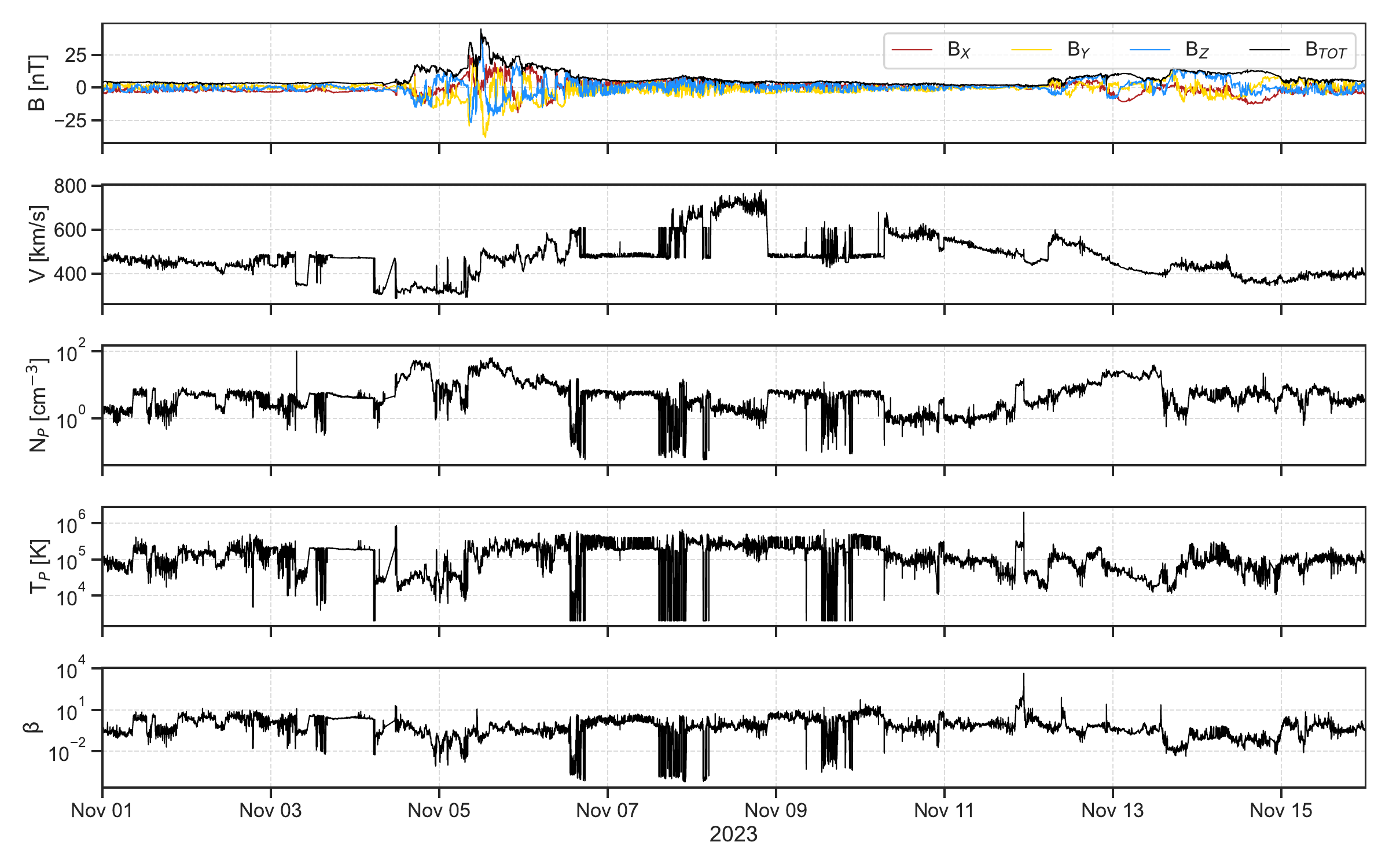}
\caption{This figure shows the real-time solar wind data as available via NOAA SWPC. The top panel shows the total magnetic field strength $|B|$, along with the vector components $B_X$, $B_Y$, and $B_Z$ in Geocentric Solar Magnetic coordinates. The remaining panels show from top to bottom: the bulk velocity $V$, the proton density $N_P$, proton temperature $T_P$, and plasma $\beta$.}
\label{fig:rtsw_noaa}
\end{center}
\end{figure}

To address this limitation, we rely on the NOAA Real-Time Solar Wind (RTSW) data set \cite{zwickl_noaa_1998}, which is made available by the Space Weather Prediction Center (NOAA SWPC) and accessible at \url{https://www.spaceweather.gov/products/real-time-solar-wind}, as shown in Figure~\ref{fig:rtsw_noaa}. This archived data set comprises measurements from various spacecraft located upstream of Earth, typically at the L1 Lagrange point, and tracked by the RTSW Network of ground stations. Data is available from 1998 onward, with the NOAA/DSCOVR satellite \cite{DSCOVR} serving as the primary operational RTSW spacecraft since July 2016, replacing the NASA/ACE spacecraft \cite{Chiu1998_ace}. In the event of issues with DSCOVR, the NASA/ACE spacecraft resumes its role as the operational RTSW source, serving as a backup. During such periods, RTSW data is provided by four ACE instruments: EPAM - Energetic Ions and Electrons \cite{EPAM_Gold_Krimigis_Hawkins_Haggerty_Lohr_Fiore_Armstrong_Holland_Lanzerotti_1998}, MAG - Magnetic Field Vectors \cite{MAG_Smith_1998}, SIS - High Energy Particle Fluxes \cite{SIS_Stone_1998}, and SWEPAM - Solar Wind Ions \cite{SWEPAM_Comas_1998}. The two DSCOVR instruments for which real-time data is available are the magnetometer (MAG) and the Faraday Cup (FC) \cite{lotoaniu_validation_2022}.

To mimic a real-time operational scenario, we compile a data set by selecting data from the spacecraft designated as operational during specific periods.

The DSCOVR MAG and FC data have been validated against equivalent science-quality data from Wind and ACE. The results demonstrate strong statistical agreement in both magnetic field measurements and the velocity components relevant to this study \cite{lotoaniu_validation_2022}. Similarly, \citeA{Lugaz_2018} highlighted a sufficiently high correlation in magnetic ejecta measurements between ACE and Wind, supporting the interchangeability of these data sets when referencing event catalogs.

Beyond data validation, \citeA{Bouriat_2022} explored the usability of ACE data for machine learning applications. While challenges such as data gaps were identified, their study suggested that integrating ACE with DSCOVR observations could help mitigate missing values, making these data sets more robust for predictive modeling. A related investigation by \citeA{smith2022} assessed the suitability of near-real-time (NRT) data for space weather forecasting. Their findings indicate that while NRT data sets are valuable for real-time hazard prediction, they exhibit increased short-term variability and occasional anomalies when compared to post-processed, science-quality data. Other studies, such as \citeA{turner_solar_2023}, demonstrated that despite the lower quality of NRT data, solar wind speed forecasts are comparable to those derived from science-level data.

Despite these limitations, certain parameters in NRT data remain reliable, according to \citeA{smith2022}. For instance, solar wind velocity typically deviates by no more than $\pm10\%$ from the post-processed data. Similarly, density and temperature measurements are generally consistent with science-quality data but may display greater uncertainty. Magnetic field measurements show a comparable level of variability, with total field magnitudes typically accurate within $\pm10\%$ of their processed values. While \citeA{smith2022} suggest that some models might be able to overcome these issues when being trained directly on NRT data, others may require additional preprocessing steps.

\begin{figure}
    \centering
    \includegraphics[width=.8\linewidth]{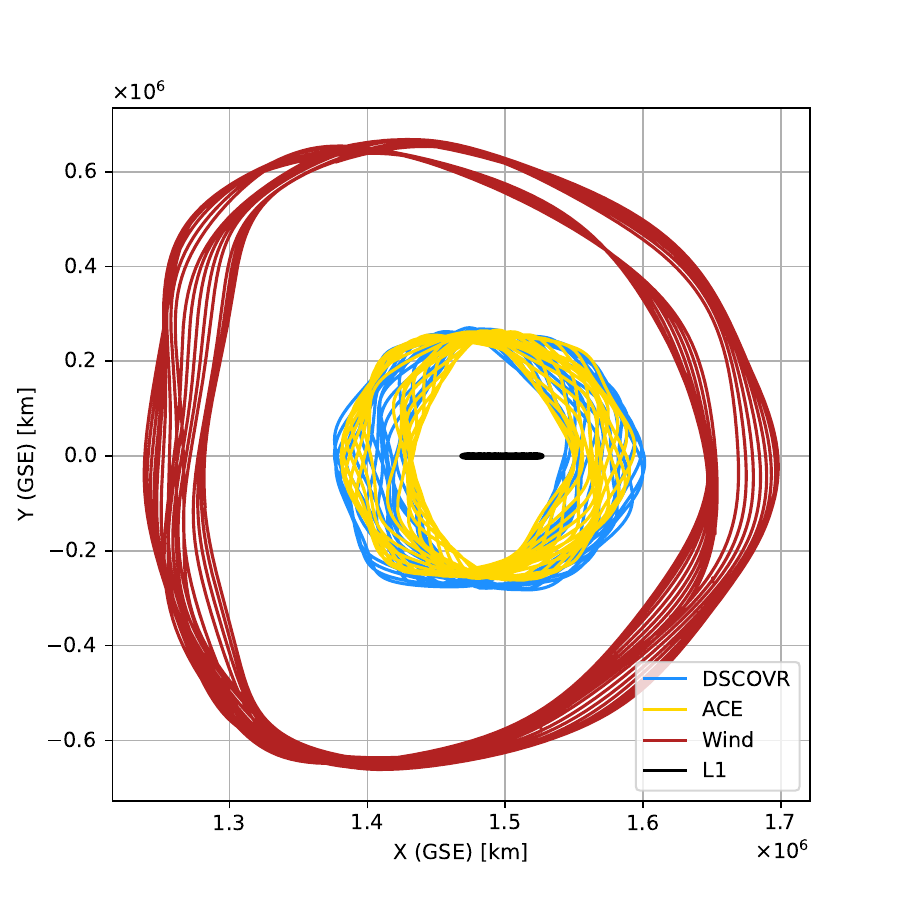}
    \caption{Spacecraft positions of Wind (blue), ACE (yellow), and DSCOVR (red) in Geocentric Solar Ecliptic coordinates from the time DSCOVR became the operational spacecraft (July 2016 onward), along with the position of the L1 point (black).}
    \label{fig:positions}
\end{figure}

Another particularity of the RTSW data is that it is not always obtained from exactly the same location. Although spacecraft such as Wind, ACE and DSCOVR are considered to be at the L1 point, they are in fact orbiting around that point. This results in slight differences in their position, as illustrated in Figure~\ref{fig:positions}. For DSCOVR and ACE, the vectorial distance between the two spacecraft ranges from approximately $1.5 \times 10^{5}$~km to $5.4 \times 10^{5}$~km, with a mean of $3.8 \times 10^{5}$~km. At these relatively short spatial scales the solar wind can be assumed to travel nearly radially. Therefore, we also compute the distance considering only the $x$ component, which yields values from about $100$~km to $1.6 \times 10^{5}$~km, with an average of $1.0 \times 10^{5}$~km. Assuming a minimum solar wind speed of $350$~\kms, this corresponds to travel times between DSCOVR and ACE of $0.0-7.6$ minutes, with an average of 4.7 minutes. At a maximum speed of $550$~\kms, the travel times range from $0.0-4.8$ minutes, with an average of 3.0 minutes. These variations illustrate that the RTSW data set is not entirely homogeneous compared to idealized single-spacecraft data, and positional offsets must be taken into account in any analysis.

In situ measurements often consist of a wide range of input parameters provided by different instruments. \citeA{nguyen_automatic_2019} demonstrated that incorporating all available parameters, including 15 channels of proton fluxes, resulted in slight improvements compared to using only magnetic field data and general plasma parameters. Yet, \citeA{pal_automatic_2024} focused solely on magnetic field measurements, and both \citeA{rudisser_automatic_2022} and \citeA{nguyen_multiclass_2025} successfully used a limited number of input variables without significantly affecting detection performance.

In this study, we focus on a limited number of input variables to ensure compatibility across different spacecraft. Following \citeA{nguyen_multiclass_2025}, we select the following parameters: the three Geocentric Solar Magnetic components of the interplanetary magnetic field ($B_X$, $B_Y$, $B_Z$), their total magnitude ($B$), the proton density ($N_p$), the proton temperature ($T_p$), the bulk solar wind speed ($V$), and the plasma beta ($\beta$). This results in a total of six independent parameters, while $B$ is derived from the individual magnetic field components and $\beta$ is calculated using $N_p$, $T_p$, and $B$. This choice is guided by both previous studies and by practical considerations regarding which measurements are consistently available across spacecraft. While a detailed feature importance or saliency analysis is beyond the scope of this work, such studies could be valuable in the future to further refine the input parameters, for example, in the context of model simplification or deployment on resource-constrained platforms.

\begin{figure}
\begin{center}
\includegraphics[width=\textwidth]{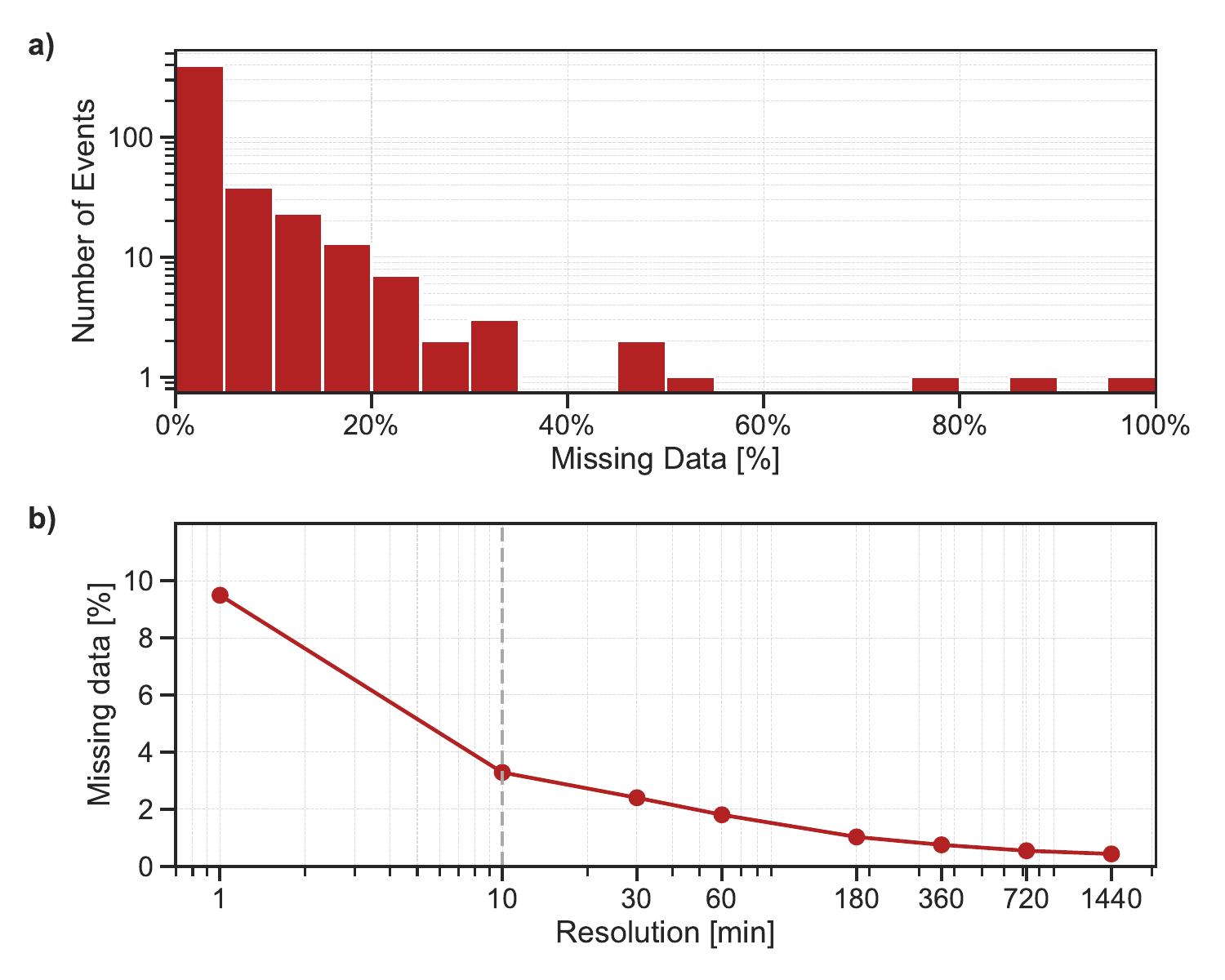}
\caption{Overview of missing data characteristics in the real-time solar wind (RTSW) data set. (a) Histogram of missing data percentage per interplanetary coronal mass ejection event for the RTSW data set. (b) Percentage of missing data depending on the chosen resolution for both the RTSW data set. The dashed vertical gray line indicates the resolution we opted for in this study (10~min).}
\label{fig:data-combined}
\end{center}
\end{figure}

Figure~\ref{fig:data-combined}a illustrates the percentage of missing data per ICME event for the RTSW data set. As expected, the RTSW data set exhibits a significant proportion of missing data. Despite these limitations, we train on the RTSW data set to enable the model to adapt to and learn from lower-quality data. Our tests have shown that this strategy enhances the model´s robustness and overall performance compared to training on a science-quality data set, such as the OMNI data set.

The complete RTSW data set comprises approximately $1.23 \times 10^{7}$ data points between 1998-02-16 00:00:00 and 2024-12-17 14:20:00, with a resolution of $1$ minute and about $9.49\%$ missing values. To address this, \citeA{nguyen_automatic_2019}, \citeA{rudisser_automatic_2022} and \citeA{nguyen_multiclass_2025} resampled all in situ data to a $10$-min resolution. Figure~\ref{fig:data-combined}b shows the percentage of missing data, depending on the chosen resolution. For this study, we likewise adopt a $10$-min resolution to balance minimizing data gaps with maintaining sufficient temporal resolution for early warning and detection. This choice is further supported by the intrinsic uncertainties arising from the different locations of the spacecraft. Additionally, this approach inherently accounts for uncertainties in event boundaries and mitigates the impact of short-term data gaps. Importantly, visual inspection confirmed that typical interplanetary shocks, which are one of the main indicators of a CME arrival \cite{salman_2018}, remain clearly identifiable and are not significantly smoothed out. This resampling strategy effectively reduced the proportion of missing values to approximately $3.28\%$ in the RTSW data set.

To further improve data continuity, we applied linear interpolation for gaps shorter than 6 hr, reducing the proportion of missing values to $<1$~\%. This approach may in some cases smooth out clear CME signatures, particularly if gaps occur near a shock, potentially complicating detection and introducing errors \cite{wang_2025}. Nevertheless, a decision on how to handle missing data is unavoidable, and linear interpolation is a widely used method. A systematic assessment of how interpolation affects CME detection, as well as a comparison with alternative gap-filling methods, would be valuable for future studies to explore.

As proposed in \citeA{rudisser_automatic_2022}, a sliding window method was applied to segment the time series and associated labels into fixed-length windows of $1024$ timesteps. At the chosen resolution of $10$ minutes, each window corresponds to a little over $7$ days. This window size is sufficient to capture even large CMEs or interacting structures in their entirety while providing enough temporal context on the background solar wind to reduce confusion with SIRs or other structures. A stride of $1$ was used to simulate a real-time operational environment. Finally, any windows that still contain missing values are discarded. This resulted in approximately $1.29 \times 10^{6}$ samples for the RTSW data set.

We normalize and scale the data such that each feature has a mean of $0$ and a standard deviation of $1$. To preserve the relative magnitudes of the magnetic field components, the four magnetic field values ($B_X$, $B_Y$, $B_Z$, $|B|$) are treated as a single feature during the scaling process.

\subsection{Event Catalogs}

As extensively discussed in prior studies, ICME catalogs often differ in their identification criteria, leading to variations in both the number of recorded events and their reported start and end times. For this study, we use the HELIO4CAST ICME catalog, which is based on in situ magnetic field and plasma measurements from multiple spacecraft in the heliosphere. Alternative catalogs were considered but deemed unsuitable for this study. The ACE catalog of \citeA{richardson_2010} provides event boundaries to the nearest hour, which is insufficient for our analysis given the 10-min data resolution. We also deliberately chose not to use the OMNI catalog used in \citeA{nguyen_multiclass_2025} as OMNI data involves propagated and merged measurements from multiple spacecraft, which introduces additional uncertainties and time shifts. The HELIO4CAST ICME catalog, described in \citeA{moestl_2017, moestl2020_icmerate}, consolidates entries from several existing catalogs, including the Wind-based catalog of \citeA{nieves_chinchilla_2018}, and is regularly updated. We use version 2.3 \cite{moestl2025_catalog} covering the period 1995-2024.

Restricting the data set to events observed at the L1 point (Wind) and excluding those with remaining data gaps, the catalog yields a total of 445 events for the time span considered in this study. As discussed in Section~\ref{sec:insitu}, there is a non-negligible difference in positions for different spacecraft at L1, which may influence the timing of in situ detections. Figure~\ref{fig:positions} illustrates that Wind follows a substantially larger orbit around L1 compared to DSCOVR and ACE. To assess whether the Wind-based catalog is nevertheless suitable for analyses using RTSW data, we carried out the following consistency analysis.

For each ICME listed in the catalog, we compared the positions of Wind and the corresponding RTSW spacecraft that was operational at the start time of the event. Based on the positional offset in the $x$-dimension (see Section~\ref{sec:insitu}) and the mean ICME speed reported in the HELIO4CAST catalog, we estimated the expected temporal shift between detections. The resulting offsets range from $0.3-14.5$ minutes, with an average of 6.5 min. Given that our study relies on in situ data resampled to a temporal resolution of 10 min, these differences are well below the effective sampling interval. We therefore conclude that the Wind-based ICME catalog can be used reliably for the purposes of this study.

In our approach, we aim to detect the entire ICME structure, including the sheath region when present, rather than restricting the detection to the MO alone. Across the full HELIO4CAST catalog time span, 533 events were observed at Wind, 407 of which include a sheath region. The mean ICME duration is 28.6 hr, with the MO averaging 21.9 hr and the sheath 8.8 hr. This choice is consistent with \citeA{nguyen_multiclass_2025} and reflects the practical and scientific importance of the sheath, which is often geoeffective and provides valuable early indicators for forecasting key ICME parameters \cite{riley_2023}.

\subsection{Generation of Labels}\label{sec:labels}

To generate the labels for the time series segmentation task, we process the catalog, following the approach described in \citeA{rudisser_automatic_2022}. Specifically, we create a binary time series where each timestep is labeled as 1 if it falls within an event and 0 otherwise. In future work, this could be adapted to assigning different values to the sheath and the flux rope part of the ICME. The start and end times of the events have been rounded to the chosen resolution of $10$ min.

Windows were classified as positive if the last timestep fell within an ICME event and was therefore labeled as $1$. An example pair of input and output is shown in Figure~\ref{fig:modelarchitecture}.

Using this criterion, the proportion of positive samples, corresponding to having the last timestep in the window labeled as ICME, was calculated to be $5.89\%$ for the RTSW data set.

\section{Framework and Methodology}\label{sec:framework}

In this section, we introduce the setup of our framework, ARCANE (Automatic Real-Time deteCtion ANd forEcast), along with its early detection module. ARCANE serves as a highly modular and adaptable machine learning framework created to address the complexities of time series event detection tasks. Its primary goal is to streamline workflows by offering integrated modules and tools for data preprocessing, model training, testing, evaluation and visualization. 

The framework is built on Hydra \cite{Yadan2019Hydra}, which provides a flexible and modular setup, making it easy to configure and manage experiments. The configurable components are organized into eight main categories: Datasets, Boundaries, Callbacks, Collates, Models, Modules, Samplers, and Schedulers. Each module can be adjusted directly through configuration files, allowing for quick modification of setups without altering the core code.

These modules integrate with available scripts, which handle tasks such as training, testing, analysis and prediction. The framework also includes routines specifically designed to download and process the RTSW data.

\subsection{Model Architecture}

The used model is a modified ResUNet++ architecture, as described in \citeA{rudisser_automatic_2022}, which achieved state-of-the-art performance for the automatic detection of ICMEs. \citeA{nguyen_multiclass_2025} compared this architecture to their YOLO-based approach, ``SPODIfY," finding comparable results between the two methods. The main difference lies in the output representation: while YOLO-based models rely on bounding boxes to localize events, ResUNet++ performs pointwise segmentation. Since our focus lies on early detection in streaming data, in which events unfold progressively and are not fully visible from the start, segmentation proves more effective than bounding boxes, which may struggle to capture the gradual onset and the temporal and spatial complexity of ICME signatures. Initial tests confirmed that ResUNet++ is better suited for our goals, demonstrating superior performance in predicting event boundaries at an early stage.

\begin{figure}
\begin{center}
\includegraphics[width=\textwidth]{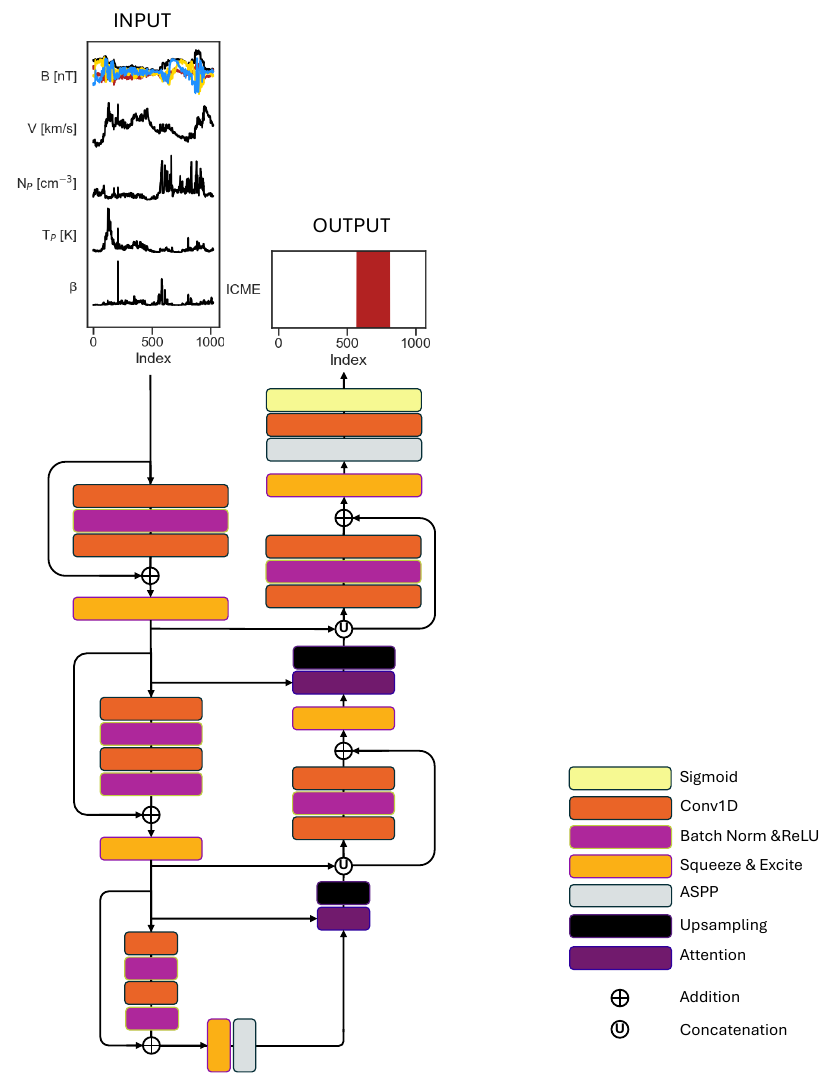}
\caption{Schematic representation of the used model, adapted from \citeA{rudisser_automatic_2022}. A complete description of the used layers and components can be found in \citeA{rudisser_automatic_2022}. Additionally, we show a sample from the data set, containing an example input and output. The input consists of 8 variables. From top to bottom: Magnetic field $B$ plus components $B_X$, $B_Y$, $B_Z$, bulk velocity $V$, proton density $N_P$, proton temperature $T_P$ and plasma $\beta$. The output is a segmented time series that consists of $0$ (white) and $1$ (red), indicating whether a given time step corresponds to an interplanetary coronal mass ejection event.}
\label{fig:modelarchitecture}
\end{center}
\end{figure}

While the original implementation in \citeA{rudisser_automatic_2022} used 2D convolutional layers, we adapted the architecture to use 1D convolutions, reflecting the structure of our data set. Apart from this modification, the overall architecture remains unchanged and is shown in Figure~\ref{fig:modelarchitecture}. For a complete description of all the components of the model, see \citeA{rudisser_automatic_2022}.

\subsection{Training}

To address the limited data set size, we employ a nested cross-validation strategy adapted from \citeA{bernoux_2022_geomagforecast}, consisting of two loops. In the outer loop, the data set is divided into yearly folds. In each iteration, 1 year is held out as the independent test set, while the remaining years are reserved for training and validation. Importantly, the test year is never used during model development.

In the inner loop, the remaining years are split into three equal subsets. Two subsets are used for training and the third for validation, and this process is repeated three times, so that each subset serves once as the validation data. Each validation fold contains at least eight years of data and each training fold at least 16 years, providing coverage of both active and quiet solar cycle phases. For each outer loop iteration, the final prediction for the held-out test year is obtained by averaging the outputs of the three models trained in the inner loop, thereby reducing variance and enhancing stability.

This two-loop evaluation approach prevents data leakage, as test years remain completely unseen during training and validation, and frames near boundaries are excluded. Although it requires training multiple models, the method yields robust performance estimates and enables a fair assessment of model skill across different phases of the solar cycle.

During training, we minimize the Dice Loss \cite{jadon_2020}, with a smoothing factor of one, using the Adam optimizer \cite{kingma2014_adam}. The model is trained for a maximum of $100$ epochs with an initial learning rate of $10^{-3}$, which is reduced by a factor of $0.1$ if the validation loss does not improve for $3$ epochs. Early stopping halts training if no improvement is observed over $10$ consecutive epochs. To account for the class imbalance, we adopt a weighted sampling approach. The probability of drawing a sample with a positive label at the last timestep is 10 times higher than for a negative sample. We use a batch size of $128$, and training samples are shuffled after each epoch. These hyperparameters have been fine-tuned through extensive optimization.

On an NVIDIA GeForce RTX 4090, 24GB GPU, one epoch takes less than $4$ min and the model typically converges in less than 10 epochs.

\subsection{Baseline}
We implemented a simple baseline to compare our results to. \citeA{lepping_2005_automaticdetection} introduced several prediction criteria to identify ICMEs in solar wind in situ data. Their classification scheme is divided into two parts: the first part focuses on the identification of ICME characteristics on short time scales ($>$ 30 min) and the second part on the identification on longer time scales ($>8$ hr). To avoid introducing a minimal waiting time of $8$ hr and simultaneously be able to perform a pointwise comparison, we focus on the first part of their classification scheme, which is based on the following criteria: the running average of proton plasma beta must be low ($< ~ \! \! \! \beta_p \! \! \! ~ >_L \ \ \leq 0.3$), the direction of the magnetic field must change slowly (quadratic fitting of $\Theta_B$ (latitude) with $\chi^2 \leq 450$), the average of the magnetic field magnitude $< \! \! \vert B\vert \! \! >$  must be $\geq 8.0$ nT, the average proton thermal velocity $<~\! \! \vert V_{Th} \vert \! \! ~> $ must be $\leq 30$ \kms and the latitudinal difference angle of the magnetic field, $\Delta \Theta_B$ must be $\geq 45$ \dg.

To adapt these criteria for a real-time setting, we define the following thresholds:

\begin{itemize}
    \item $B_{max} \geq 8 $ nT 
    \item $\beta \leq 0.3 $
    \item $T_p \leq 4.3 \times 10^{4}$ K
\end{itemize}

The threshold for $T_p$ is derived from the equation:

\begin{equation}
      T_p = \frac{\pi m_p V_{Th}^2}{8 k_B}
\end{equation}

where $k_B$ is the Boltzmann constant and $m_p$ is the proton mass. $T_p$ is given in Kelvin, and $V_{Th}$ in \kms. The value is rounded up for simplicity. During inference, each time step in the data set is analyzed individually to determine whether it meets all the conditions required to classify it as an event.

\subsection{Postprocessing}

During inference, the original ResUNet++ model processes data in sliding windows and outputs a time series of values between zero and one for each window. To evaluate the model´s ability to detect events early, we extract the model´s prediction at the last time step of each window and stack them to obtain a time series. This extracted time series, denoted as $t_{1}$, represents the classification decision that would be made as soon as a new point in time enters the window. This approach simulates real-time classification, where decisions must be made without waiting for future data.

To systematically analyze how early the model can reliably detect an event, we extend this approach to earlier time steps within the window. Specifically, we extract predictions at progressively later time steps, denoted as $t_{2}$ through $t_{150}$. Each of these corresponds to a different waiting time $\delta$, which we define as the duration for which a point in time has been observed before being classified. Per definition, $t_{6}$ accounts for the prediction that has been made after observing a point in time for $\delta (t_{6}) = 1 $ hour, taking into account an additional data point and $t_{150}$ accounts for the prediction after $\delta (t_{150}) = 25 $ hr.

\begin{figure}
\begin{center}
\includegraphics[width=.8\textwidth]{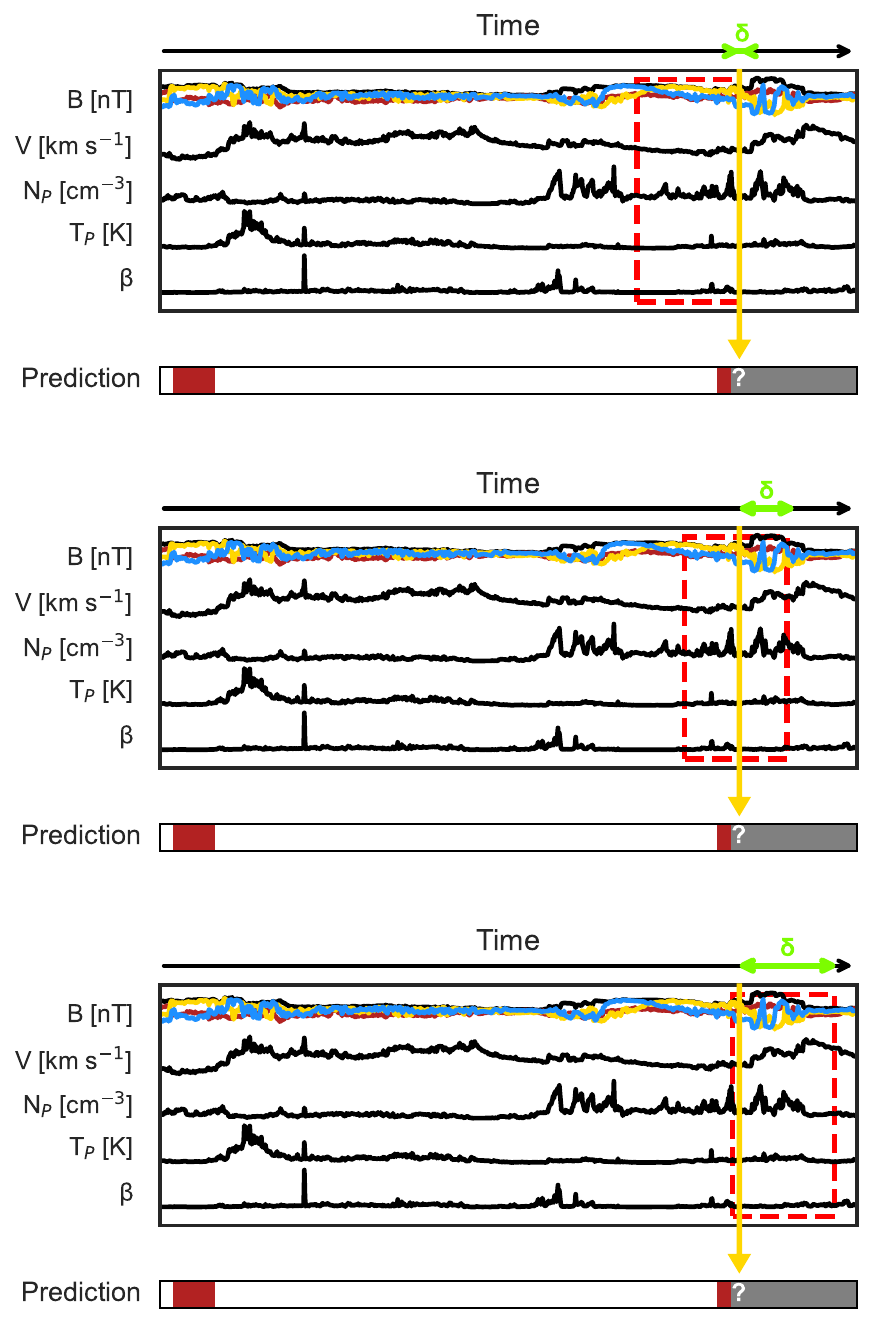}
\caption{Schematic illustration of the postprocessing method used to extract 150 different time series of predictions at varying waiting times $\delta$.
Each subplot represents a single prediction time step for a specific $\delta$ value. The red dashed rectangle indicates the sliding window of input data provided to the model. The yellow vertical arrow marks the current time step being classified as either CME or no CME. The horizontal strip below each plot labeled Prediction represents the full prediction time axis: segments before the prediction point correspond to previous predictions (red to indicate positive predictions), while the gray segment under and after the yellow arrow indicates time steps that have not yet been predicted.
The neon green double-headed arrow labeled $\delta$ shows the gap between the end of the input window and the current prediction time step, defining the waiting time: how long the model “waits” before making a prediction about a point in time.
By shifting the prediction point further from the input window (increasing $\delta$), we can assess how the model´s detection performance evolves with seeing more future data before making a prediction.}
\label{fig:postprocessing}
\end{center}
\end{figure}

This method produces 150 different time series, each representing the model´s predictions at a specific waiting time $\delta$. By analyzing these series, we can study how detection performance evolves as more data becomes available. A schematic representation of this process is shown in Figure~\ref{fig:postprocessing}.

The final postprocessing steps are identical for both the ARCANE Classifier and the Threshold Classifier baseline. To convert each time series into a list of events $E$, we apply a simple thresholding approach for a range of thresholds between $0.1$ and $1.0$. Consecutive time steps exceeding the threshold are grouped into a single event.

To ensure practical applicability, we discard events with a duration of $ \leq 10 $ minutes, as these would not be detected in a real-time setting. Additionally, events separated by $ \leq 10$  minutes are merged into a single event. The final boundaries of the detected events $E_d$ are determined by the first ($t_s(E_d)$) and last ($t_e(E_d)$) time step that exceeds the threshold.

\citeA{nguyen_multiclass_2025} used a slightly different approach only applicable in a non-real-time scenario. Their method applied a fixed threshold of $0.1$, and the detection probability of each event was computed as the mean probability within its detected boundaries. While we compared our results to theirs by replicating their postprocessing approach as a benchmark, we primarily rely on our own postprocessing method, better suited to our real-time detection requirements.

\subsection{Evaluation}\label{sec:evaluation}

In the postprocessing step, we generate a list of detected events, denoted as $E_d$, and compare it to the list of ground truth events $E_t$. For each ground truth event, we check whether it overlaps with any detected event. If an overlap is found, the ground truth event is counted as a true positive (TP). In cases where multiple detected events overlap with a single ground truth event, only the earliest overlapping detected event is assigned as a true positive; the remaining overlaps are ignored for this event. If a ground truth event does not overlap with any detected event, it is considered a false negative (FN). Conversely, any detected event that does not overlap with a ground truth event is counted as a false positive (FP).

We calculate the standard metrics Precision, Recall and F1-Score to evaluate the model´s performance:

\begin{equation}\label{eq:precision}
    \text{Precision} = \frac{\text{TP}}{\text{TP} + \text{FP}} 
\end{equation}

\begin{equation}\label{eq:recall}
    \text{Recall} = \frac{\text{TP}}{\text{TP} + \text{FN}} 
\end{equation}

\begin{equation}
      \text{F1} = 2 \times \frac{\text{Precision} \times \text{Recall}}{\text{Precision} + \text{Recall}}
\end{equation}

These metrics directly quantify the trade-off between missed events and false alarms, which is most relevant for operational ICME detection. Metrics that rely on true negatives, such as specificity or the Matthews Correlation Coefficient, are not reported here, since true negatives are not uniquely defined in our event-based evaluation framework. Periods without ICMEs could be mapped to multiple “non-event” classes depending on assumptions about event duration and windowing, making the number of true negatives highly sensitive to evaluation choices rather than model skill.

To specifically assess the model´s ability to detect events early, we introduce a new metric, ``Delay". This metric measures the time difference between the actual start time of a ground truth event $t_s(E_t)$ and the time of its first detection, where the first detection corresponds to the time step $t_s(E_d)$, at which the threshold is exceeded for the first time. To account for the model´s prediction delay, we add the time series´ waiting time parameter $\delta$ to the detected events´ start time:

\begin{equation}
      \text{Delay} = \text{max}((t_s (E_d) - t_s(E_t), 0) + \delta
\end{equation}

This definition ensures that early or perfectly timed detections result in a Delay equal to the waiting time $\delta$, while late detections are penalized by the additional time lag. The max function ensures that early detections do not lead to artificially negative or reduced Delay values, since in operational practice, an early alert is still subject to the waiting time $\delta$.

While related, $\delta$ and Delay serve different roles: $\delta$ represents the chosen observation window prior to classification, whereas Delay measures the effective time to detection by combining $\delta$ with any additional lag between the ground-truth onset and the model´s first detection.

Finally, we also report the error on start time, corresponding to the absolute value $|(t_s (E_t) - t_s(E_d)|$.

\begin{figure}
\begin{center}
\includegraphics[width=\textwidth]{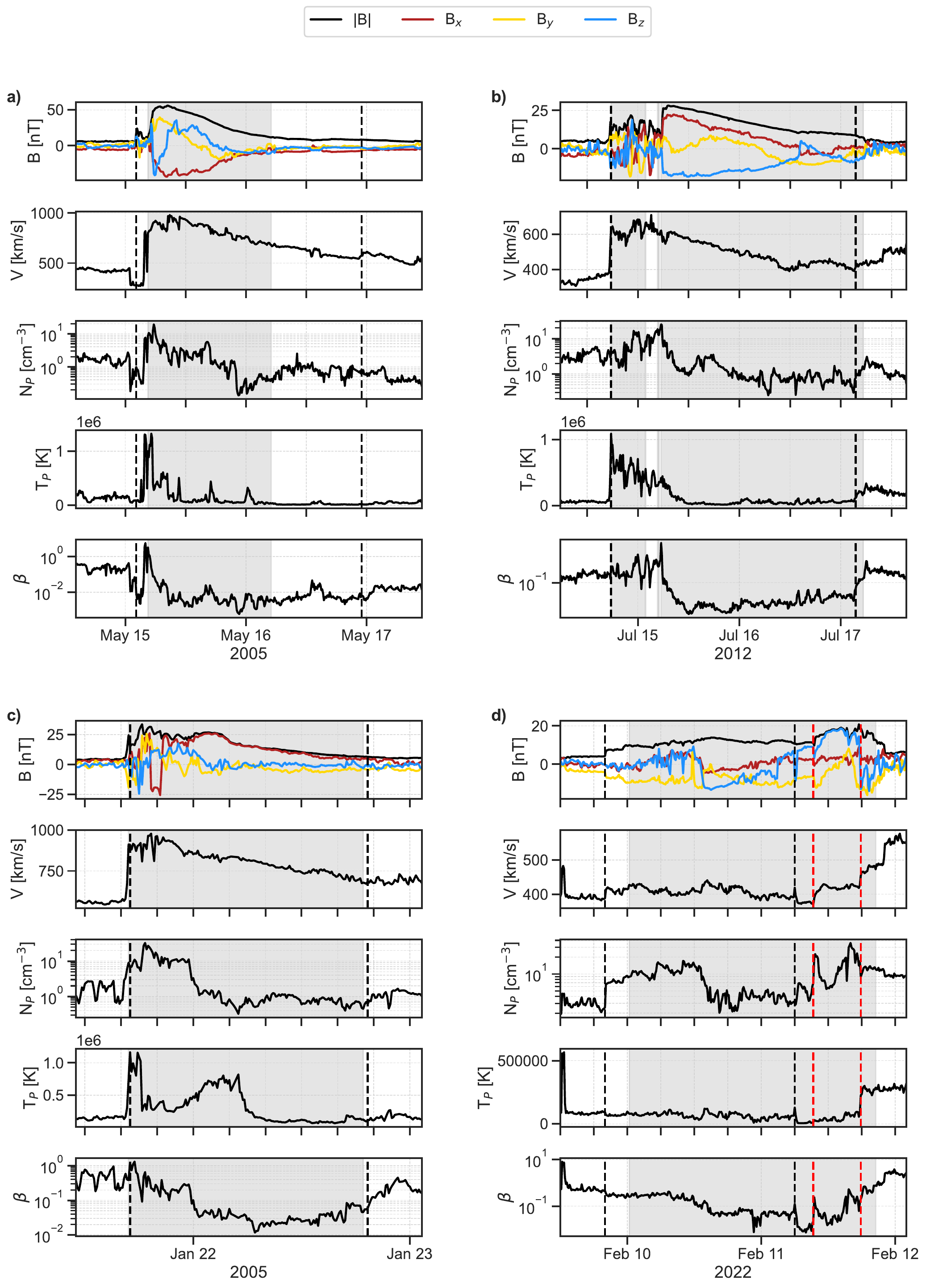}
\caption{Four examples showing both the ground truth and the prediction made by ARCANE. The total magnetic field $|B|$ along with its vector components $B_X$, $B_Y$ and $B_Z$ are displayed, with the gray shaded area corresponding to the events predicted by ARCANE and the vertical black and red lines denoting the start and end time of the respective ground truth events. (a) Correctly predicted event with a small error on the start time. (b) Single ground truth event predicted through multiple events. (c) Correctly predicted event with no error on the start time. (d) Multiple ground truth events predicted through a single event.}
\label{fig:examples}
\end{center}
\end{figure}

Figure~\ref{fig:examples} illustrates multiple scenarios in which these considerations are particularly important. Figure~\ref{fig:examples}a shows a correctly predicted event with a small error on the start time. The detected event does not extend to the trailing edge of the ICME, likely due to unclear boundary signatures in the magnetic field, density and temperature. As a result, ARCANE may have assigned a lower probability to the final part of the structure. Since Figure \ref{fig:examples} displays the binary classification after applying the threshold, this section of the ICME likely remained below the decision boundary, even though the raw probability output could have been nonzero. Figure~\ref{fig:examples}b shows a single ground truth event, for which two predicted events are detected. The second predicted event is not counted as an additional TP to avoid overestimating the overall Precision. This choice reflects the logic of a real-time detection setting, where the first prediction would have already triggered an alert. Any further predictions for the same event are considered redundant.

Figure~\ref{fig:examples}c shows a correctly predicted event with no error on the start time. Figure~\ref{fig:examples}d shows a case where two distinct ground truth events are both detected by the same predicted event. In this case, the start time error for each ground truth event is measured as the difference between the beginning of the predicted event (shaded region) and the respective true start times, denoted by the vertical black (first event) and red (second event) lines. This leads to a larger start time error for the second event. Still, the Delay for the second event is defined as $\delta$, since the contribution of the error on the start time to the Delay cannot be negative.

\subsection{Validation of Postprocessing and Evaluation}

\begin{figure}
\begin{center}
\includegraphics[width=\textwidth]{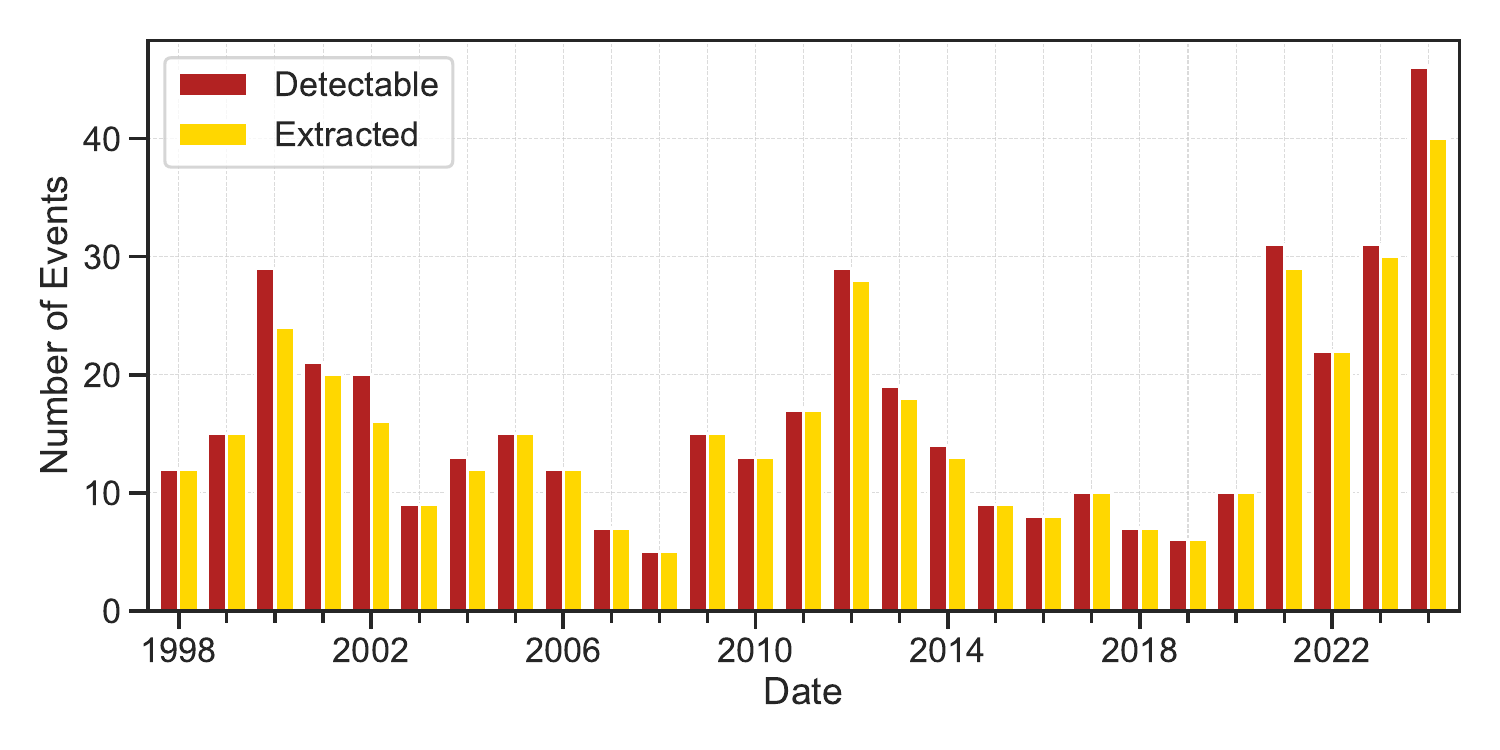}
\caption{Number of events over the entire real-time solar wind data set for both the ground truth (detectable) and the generated catalog (extracted). The dependence of the number of events on the solar cycle of $11$ years is clearly visible.}
\label{fig:event_numbers_rtsw}
\end{center}
\end{figure}

To test the validity of this approach, we attempt to regenerate the catalog from our created labels and evaluate the two catalogs against each other. This analysis is conducted to verify that the forward and backward mapping between the generated time series and the event catalog works as expected. We calculate the Precision, Recall and F1-Score for the generated catalog and compare it to the ground truth catalog. Figure~\ref{fig:event_numbers_rtsw} shows the number of events over the entire data set for both the ground truth and the generated catalog. 

As expected, the number of events in the generated catalog is lower as the approach combines nearby events into a single event. Our approach is supposed to allow for this behavior without penalizing it in the evaluation if only one of these nearby events is detected. In the case of the RTSW data set, there are $445$ events to be detected. Extracting $422$ events during the postprocessing step, the evaluation yields a perfect Precision, Recall, and F1-Score of $1.0$. As these merged events cannot be distinguished, we treat the generated catalog as our ground truth moving forward.

\section{Results}\label{sec:results}

\subsection{Detection Performance and Delay}

\begin{figure}
\begin{center}
\includegraphics[width=\textwidth]{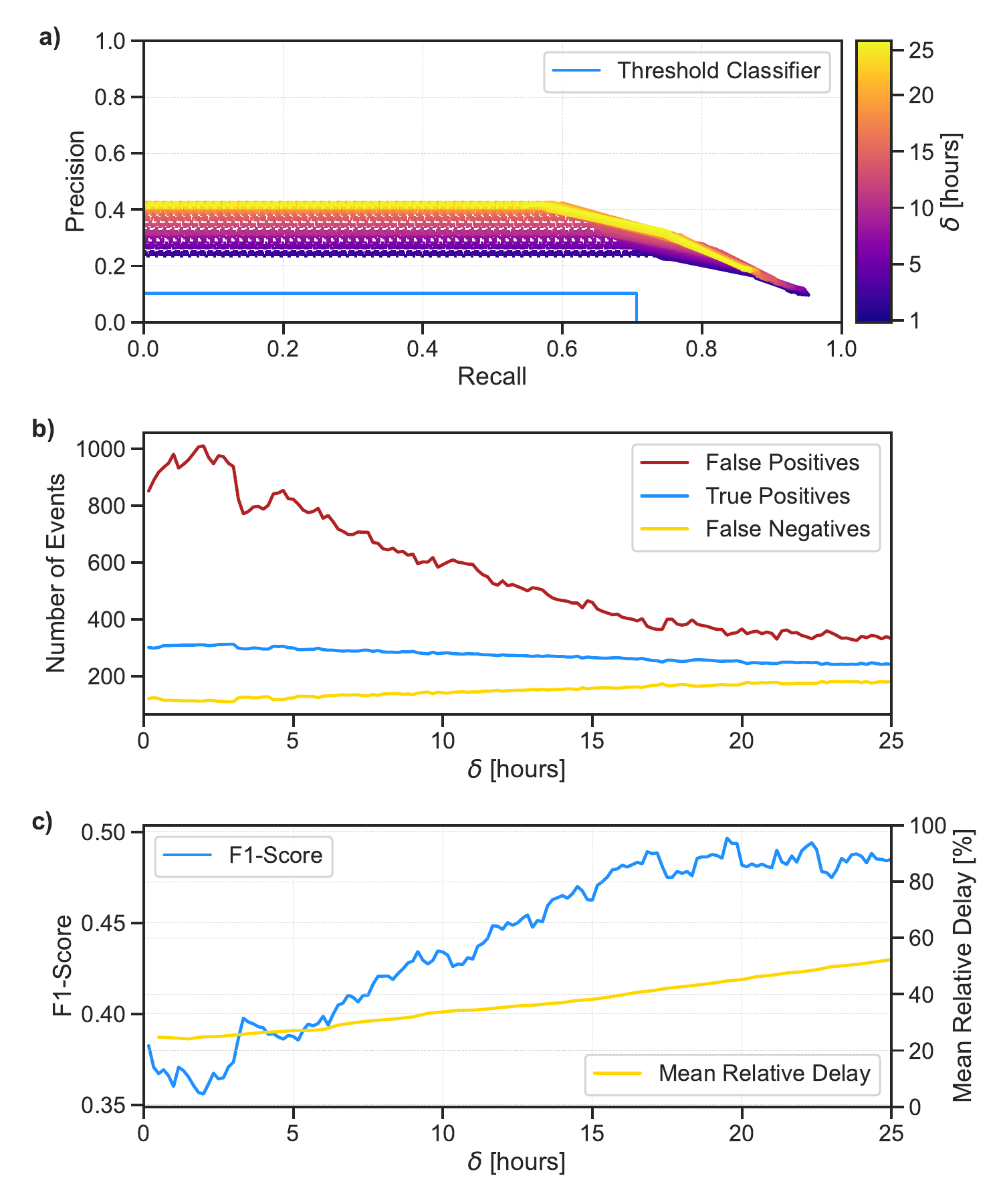}
\caption{Overview of detection performance as a function of waiting time $\delta$. (a) Precision-Recall curves for each time series. The color indicates the waiting time $\delta$ in hr. The performance Threshold Classifier baseline is indicated as a blue rectangle for comparison. (b) Number of False Positives, True Positives and False Negatives as a function of the waiting time $\delta$. (c) F1-Score and mean relative Delay in percentage of the duration as a function of the waiting time $\delta$.}
\label{fig:combined-results}
\end{center}
\end{figure}

As explained in Section~\ref{sec:evaluation}, we generate an event catalog from the model´s predictions and compare it to the ground truth catalog for each of the $150$ time series, based on different waiting times $\delta$. Precision and Recall are calculated across a range of thresholds for each time series and the resulting Precision-Recall curves are shown in Figure~\ref{fig:combined-results}a, where the color indicates the waiting time $\delta$ in hr. The curves are horizontally extended as dashed lines for better comparability. The performance of the Threshold Classifier baseline is indicated as a blue rectangle for comparison. The Threshold Classifier exhibits very low Precision, which can be attributed to a large number of false positives. This behavior is expected, as at a 10-minute resolution even relatively small-scale fluctuations may exceed the threshold and be incorrectly classified as CME signatures.

As anticipated, longer waiting times result in higher Precision values, as the model is exposed to a larger portion of the event by that point in time. However, this improvement appears to be relatively modest after the initial hr have passed. Interestingly, shorter waiting times yield slightly higher Recall. This observation aligns with the model´s tendency to generate optimistic alerts based on small variations when it has seen only a fraction of an event. Over time, as more data becomes available, the model refines its predictions, reducing the number of false positives and thereby increasing Precision.

Figure~\ref{fig:combined-results}b shows the number of False Positives, True Positives, and False Negatives as a function of the waiting time $\delta$. The very high number of False Positives at short waiting times, which then decreases substantially with increasing $\delta$, explains the high Recall values observed for low waiting times in Figure~\ref{fig:combined-results}a.

Additionally, we compute the maximum F1-Score for each waiting time $\delta$ and present it as a function of $\delta$ in Figure~\ref{fig:combined-results}c. This figure highlights the clear dependence of model performance on the waiting time. Although waiting for extended periods is impractical in a real-time scenario, the performance only approaches its maximum after around 16 hr. Comparing this timescale to typical sheath durations ($\sim$12 hr at ACE \cite{janvier_2019_generic} or 8.8 hr on average at Wind in the HELIO4CAST catalog) may indicate that full information on the sheath as well as parts of the MO are necessary to reliably distinguish between CME-driven sheaths and driverless or SIR-driven sheaths. Since the primary interest lies in detecting the MO, however, the model can still provide useful early warnings before the maximum performance is reached, which is important for operational contexts.

We further evaluate the model´s ability to detect events early by calculating the delay parameter for each event and waiting time. The mean relative delay for each waiting time is shown in Figure~\ref{fig:combined-results}c and exhibits a linear relationship, as expected. The waiting time has the most significant impact on the delay, highlighting the importance of early event detection. At the same time, we observe that the error made on the start time of the event stays more or less constant and is only slightly refined as the model sees more of the event. While the F1-Score increases substantially within the first few hr, the mean delay also rises, which is crucial to minimize when it comes to operational space weather monitoring.

\begin{table}
\caption{Precision, Recall, F1-Score, Absolute Mean Error (AME) on the Start Time, Relative Mean Error (RME) on the Start Time, Mean Delay (MD) and Relative Mean Delay (RMD) for Different Waiting Times of the ARCANE Classifier and the Threshold Classifier Baseline. Best values are highlighted in bold.}
\centering
\begin{tabular}{l c c c c c c c}
\hline
      & Prec. & Rec. & F1 & AME [h] & RME [\%] & MD [h] & RMD [\%] \\
\hline
      ARCANE ($\delta = 0.5$h)   & 0.25 & 0.71 & 0.37 & 7.5 & 33.2 & \textbf{6.8} & \textbf{24.1} \\
      ARCANE ($\delta = 3.0$h)   & 0.25 & \textbf{0.74} & 0.37 & 6.6 & 29.4 & 8.6 & 31.9 \\
      ARCANE ($\delta = 8.0$h)   & 0.31 & 0.68 & 0.42 & 6.0 & 27.9  & 12.8 & 50.8 \\
      ARCANE ($\delta = 16.0$h)  & \textbf{0.39} & 0.62 & \textbf{0.48} & \textbf{5.5}  & \textbf{27.6} & 19.9 & 82.9 \\
      Threshold Classifier  & 0.10 & 0.71 & 0.18  & 10.0 & 32.0 & 9.9 & 31.7  \\
\hline
\label{tab:eventwise_results}
\end{tabular}
\end{table}

We summarize the maximum Precision, Recall, and F1-Score for four different waiting times of the ARCANE Classifier and the Threshold Classifier baseline in Table~\ref{tab:eventwise_results}. Additionally, we show the absolute and relative mean error on the start time. To quantify the model´s ability to detect events early, we calculate both the mean and relative mean of the Delay parameter for the four different waiting times., as well as for the Threshold classifier baseline.

\begin{figure}
\begin{center}
\includegraphics[width=\textwidth]{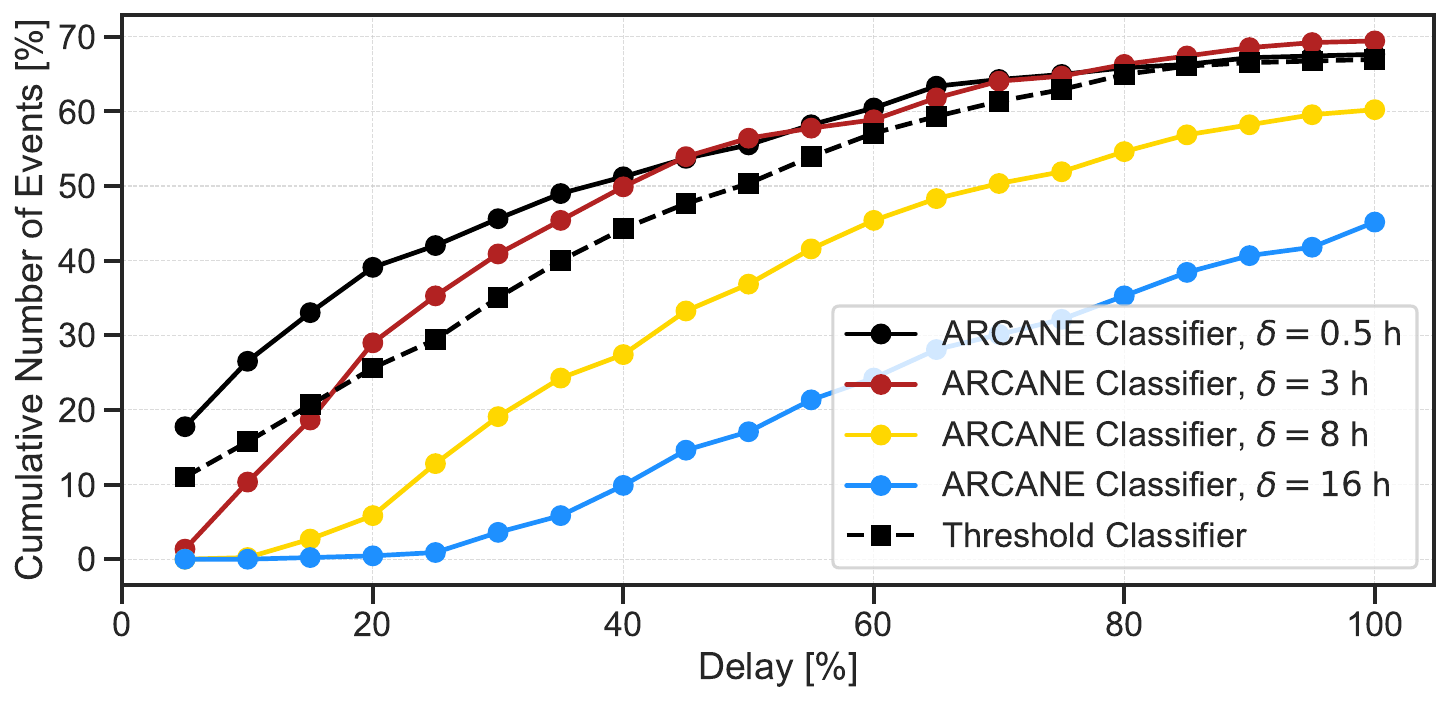}
\caption{Cumulative distribution of the Delay values for the ARCANE Classifier at different waiting times and the Threshold Classifier baseline. The Delay is given in percentage of the event duration.}
\label{fig:delay_distribution}
\end{center}
\end{figure}

The cumulative distribution of these delays for different waiting times is shown in Figure~\ref{fig:delay_distribution}. As expected, the cumulative number of detected events rises steeply at lower delay values before leveling off at higher delays. 

At a waiting time of $\delta = 0.5$h, the ARCANE classifier outperforms the baseline across all delay values, demonstrating its ability to detect events earlier. In contrast, at higher waiting times, the ARCANE classifier exhibits larger delay values than the Threshold classifier. Nevertheless, this increased delay is accompanied by a significantly higher F1-Score, as shown earlier, indicating a trade-off between detection performance and timeliness. It should also be noted that the Threshold Classifier is designed to detect MOs rather than the entire CME. By definition, this leads to an inherent delay compared to ARCANE, since the sheath region is not part of its detection target.

It is important to note that these delay values were calculated using a threshold optimized for maximizing the F1-Score. Adjusting the threshold further impacts the delay, allowing for a trade-off between detection performance and timeliness. By lowering the threshold of the ARCANE classifier, one could prioritize earlier event detection at the expense of a reduced F1-Score, offering additional flexibility depending on operational requirements.

\subsection{Analysis of Key Parameters}

\begin{figure}
\begin{center}
\includegraphics[width=\textwidth]{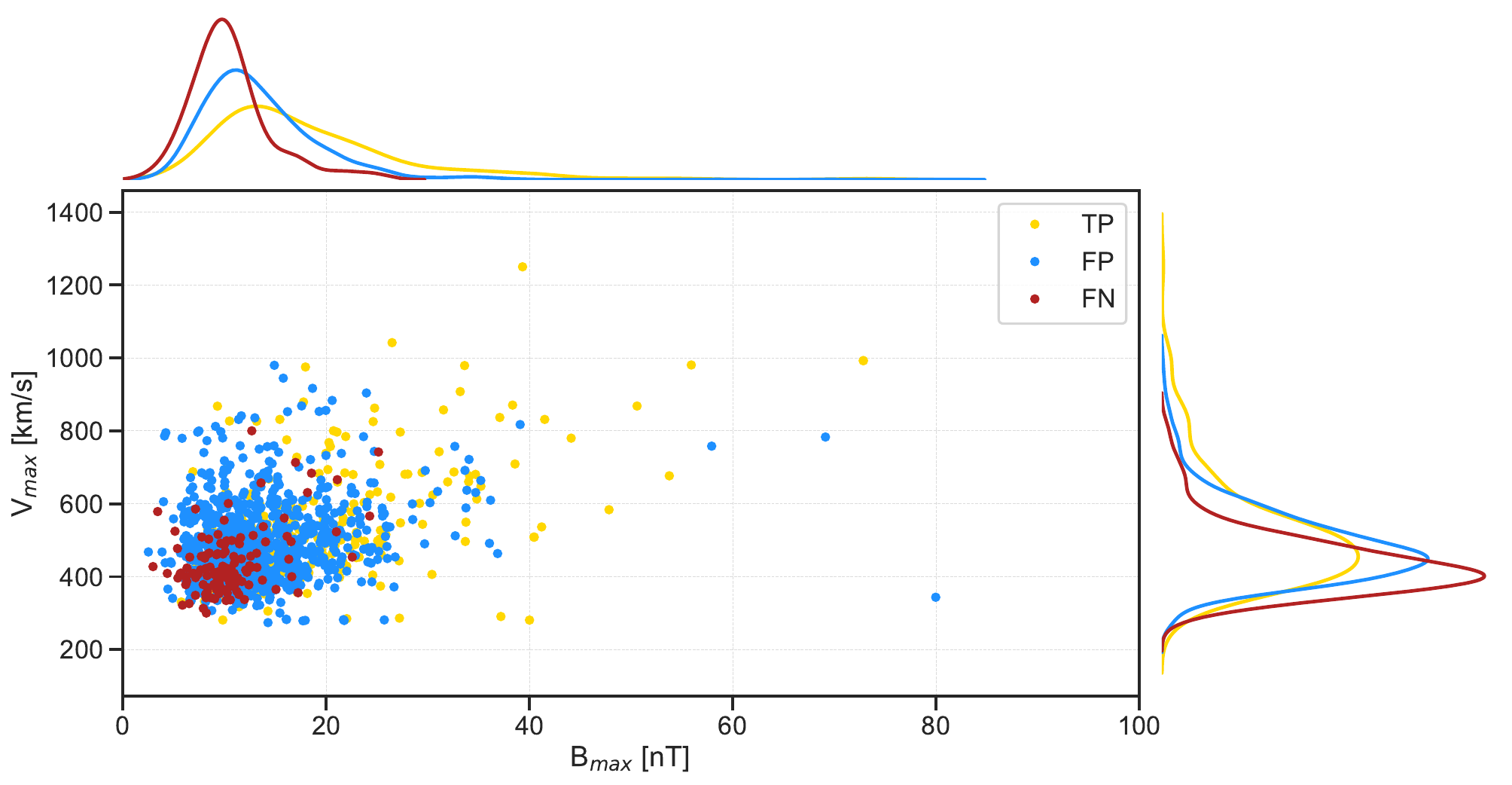}
\caption{Maximum value of $|B|$ against maximum value of $V$ for true positive, false positive and false negative events. Additionally, the kernel density estimation (KDE) for each group of events is given.}
\label{fig:key_parameters_tp}
\end{center}
\end{figure}

To better understand the characteristics of theTP, FP, and FN events at a waiting time of $0.5$h, we analyze key event parameters and their interdependencies. Specifically, Figure~\ref{fig:key_parameters_tp} visualizes the relationship between the maximum value of $|B|$ and the maximum value of $V$ for each event, alongside their kernel density estimates.

Our analysis reveals that events with high peak $|B|$ values are consistently detected, highlighting the model´s ability to effectively capture strong magnetic field structures. Since ICMEs with large magnetic field strengths can result in stronger geomagnetic storms, detecting these events is particularly important. Notably, most FN events exhibit peak $|B|$ values below 20 nT, with a significant fraction below 10 nT. Similarly, undetected events tend to have lower velocities, predominantly below 600 km~s$^{-1}$. Given that weaker magnetic field strengths and lower velocities are generally associated with less impactful ICMEs, the model´s focus on high-impact events aligns well with operational forecasting priorities.

These results indicate that the model successfully prioritizes detecting strong events while maintaining a balance in minimizing false negatives among weaker events.

\section{Conclusion and Outlook}\label{sec:conclusion}

In this study, we present ARCANE, a novel machine learning framework designed for the real-time detection and forecasting of ICMEs using in situ solar wind data. Our results demonstrate that ARCANE outperforms traditional threshold-based detection methods in both Precision and timeliness, even when applied to RTSW data obtained at the Sun–Earth L1 point. Moreover, its modular design allows for the seamless integration of additional data sources. For instance, sub-L1 monitors such as Solar Orbiter \cite{laker_2024, davies_2025_realtimepred} or STEREO-A when it crossed the Sun–Earth line during 2023--2024 \cite{lugaz_2024, weiler_2025} could provide earlier warnings or serve as auxiliary inputs to enhance detection performance. In principle, ARCANE´s detection algorithm could even be deployed onboard spacecraft to trigger high-resolution observations of magnetic fields or particle fluxes, improving space weather monitoring capabilities. Additionally, coupling ARCANE with CME arrival time models, such as ELEvo \cite{moestl2015elevo}, could further refine in situ detection probabilities.

A key advantage of ARCANE is its adaptability to different operational needs. Since the framework relies on a classification threshold to optimize detection performance, this threshold can be adjusted depending on operational requirements. Lowering the threshold prioritizes early detection, allowing for faster warnings at the cost of reduced Precision, whereas a higher threshold improves accuracy but may delay event identification. This flexibility ensures that ARCANE can be fine-tuned for specific space weather monitoring objectives, balancing timeliness and reliability based on the needs of forecasters. 

Nevertheless, further work is needed to optimize how ARCANE´s outputs are leveraged in practice. In this study, we deliberately focused on isolating the impact of the waiting time parameter on detection performance, without exploring how best to combine predictions across different waiting times in a real-time setting. In particular, we did not address how to dynamically adjust classification thresholds as new data becomes available. Future work could focus on fine-tuning these parameters to fully exploit ARCANE´s early warning capabilities while simultaneously improving the accuracy and consistency of automated event catalog generation.

One of the main challenges in advancing ICME detection lies in the limitations of existing event catalogs. While ARCANE effectively identifies high-impact events, its ability to differentiate between high- and low-severity events is constrained by the lack of severity labels in current data sets. The development of enhanced event catalogs, which is an ongoing effort in the community, including detailed severity classifications could significantly improve the framework´s performance. The computational efficiency of the framework ensures that retraining with improved catalogs or exploring the impact of alternative catalog types can be achieved at a low cost. This retraining efficiency also facilitates future studies on feature importance. Such analyses will be particularly valuable as not all spacecraft provide the same set of measurements, and they could further help to identify redundant inputs and refine the parameter set used by ARCANE.

An ideal data set for advancing detection capabilities would consist of fully segmented time series that differentiate between all possible solar wind structures. Such a data set would not only distinguish between ICME components-like shocks, sheaths, and flux ropes-but also include features like heliospheric current sheets, SIRs, and co-rotating interaction regions. This level of granularity would provide ARCANE with a comprehensive training resource, enabling it to learn the nuanced signatures of various solar wind structures and significantly enhance its overall accuracy and utility.

An alternative approach to overcoming current data set limitations is the simultaneous prediction of key ICME parameters, such as minimum $B_Z$, maximum $B$, and duration. Predicting these parameters alongside detection would allow the framework to distinguish between high- and low-severity events, enhancing its operational utility for real-time forecasting.

A key aspect of early ICME detection is its connection to the broader field of Early Time Series Classification (ETSC), which seeks to classify time series data as early as possible while balancing the trade-off between prediction accuracy and timeliness \cite{Dachraoui_2015_earlyclass, Zafar_2021_earlyclass, Bilski_2023_calimera}. ETSC solutions often employ adaptive stopping rules to optimize this trade-off in real-time applications. While the potential for ETSC in space weather prediction is substantial, most ICME detection methods have yet to fully explore their early detection capabilities or systematically evaluate the impact of different waiting times on performance. Additionally, many ETSC approaches assume the availability of near-perfect classifiers when the full data set is accessible, a condition not yet met in ICME detection \cite{nguyen_automatic_2019, rudisser_automatic_2022, pal_automatic_2024, nguyen_multiclass_2025}. Nevertheless, by incorporating ETSC principles, ARCANE could further refine its real-time decision-making strategies, enabling earlier and more reliable warnings.

A critical next step for ARCANE´s development is the integration of physical models into its detection pipeline. By combining machine learning with physics-based models, the framework could provide more comprehensive forecasts, including detailed insights into CME propagation and geoeffectiveness. Additionally, incorporating ensemble methods, where multiple models with different architectures contribute to predictions, could improve both the robustness and interpretability of ARCANE. This would enhance detection reliability and provide deeper insights into the key data features influencing ICME identification.

Compared to other machine learning approaches for ICME detection \cite{nguyen_automatic_2019, nguyen_multiclass_2025}, the F1-Scores achieved with ARCANE appear relatively modest. However, this is largely a consequence of the catalog employed. As shown in \citeA{rudisser_automatic_2022}, using the catalog of \citeA{moestl2020_icmerate} leads to considerably lower scores for Wind data compared to studies based on other catalogs. This is expected, as detection performance depends strongly on the reference catalog. In fact, \citeA{rudisser_automatic_2022} demonstrated that many of the apparent false positives actually resemble CMEs, suggesting that the catalog´s strict criteria may exclude some events. By contrast, the catalog used in \citeA{nguyen_multiclass_2025} applies looser criteria and even incorporates additional CMEs identified by their machine learning algorithm, naturally resulting in higher reported scores. We could not directly use this catalog, as it is based on OMNI data, which cannot easily be translated to real-time data. Our F1-Scores are also somewhat lower than those reported in \citeA{rudisser_automatic_2022}, but this can be explained by two factors: first, we are explicitly working with real-time data rather than higher-quality science data, and second, we systematically evaluate early detection performance, whereas previous studies focused on classification once the full structure was already observed.

Beyond these catalog-related considerations, an important physical insight from this work is that the maximum F1-score is only reached after waiting times longer than the typical sheath duration. This suggests that full knowledge of the sheath is necessary to reliably distinguish between CME-driven sheaths and driverless or SIR-driven sheaths, highlighting a fundamental limitation for early event classification.

Despite these constraints, ARCANE represents a significant advancement in the operational detection and forecasting of ICMEs. Its ability to process real-time data, flexible modular setup, and computational efficiency make it a strong candidate for ongoing improvements, especially as improved catalogs become available, additional spacecraft provide earlier in situ measurements, or future studies refine how ARCANE´s probabilistic outputs and uncertainties are interpreted and used operationally. Furthermore, ARCANE´s predictions can be combined with physics-based propagation models, such as ELEvo \cite{moestl2015elevo}, to constrain detection windows when CMEs are expected to arrive at Earth. Future developments may also integrate ensemble predictions, probabilistic outputs, and physical constraints to better quantify uncertainty and enhance practical decision-making. A transparent understanding of these limitations does not diminish the utility of ARCANE but instead provides a realistic roadmap towards a fully operational early-warning system.

%
%

\section*{Open Research}\label{sec:data_availability}

Developed specifically for operational space weather applications, ARCANE is already deployed as a prototype operational model at the Austrian Space Weather Office, accessible at \url{https://helioforecast.space/}.

To ensure reproducibility and to facilitate future research comparisons with our findings, we have made the source code and related data publicly available as follows:

The latest version of the ARCANE framework is accessible on GitHub at \url{https://github.com/hruedisser/arcane}. To enable the community to replicate and build on our work, the source code for generating the figures in this study is provided as Jupyter notebooks, available at \url{https://github.com/hruedisser/arcane/tree/main/scripts/notebooks}.

The in situ solar wind data used in this study was originally obtained from \url{http://services.swpc.noaa.gov/text/rtsw/data/}. This data, along with the ICME catalog and trained models is preserved at \citeA{ruedisser2025_data} for long-term accessibility and reference.

\acknowledgments

H.T.~R., E.E.~D., and C.~M. are supported by ERC grant (HELIO4CAST, 10.3030/101042188). Funded by the European Union. Views and opinions expressed are however those of the author(s) only and do not necessarily reflect those of the European Union or the European Research Council Executive Agency. Neither the European Union nor the granting authority can be held responsible for them.

This research was funded in whole or in part by the Austrian Science Fund (FWF) [10.55776/P36093] (J.~L.L.). For open access purposes, the author has applied a CC BY public copyright license to any author-accepted manuscript version arising from this submission.

The research leading to these results is part of ONERA Forecasting Ionosphere and Radiation belts Short Time Scale disturbances with extended horizon (FIRSTS) internal project (G.N.).

We have benefited from the availability of the NOAA RTSW data, and thus would like to thank the instrument teams and data archives for their data distribution efforts. 

During the preparation of this work, the authors used ChatGPT (OpenAI) to partly assist with improving grammar, language, and readability of the manuscript. The tool was not used for data analysis, result generation, or drawing scientific conclusions. All scientific content, interpretations, and conclusions are solely those of the authors, who take full responsibility for the content of the published article.

\section*{Conflict of Interest Statement}

The authors have no conflicts of interest to disclose.

\newpage

\bibliography{bibliography.bib}

@Misc{Yadan2019Hydra,
  author =       {Omry Yadan},
  title =        {Hydra - A framework for elegantly configuring complex applications},
  howpublished = {Github},
  year =         {2019},
  url =          {https://github.com/facebookresearch/hydra}
}

@Inbook{Chiu1998_ace,
    author="Chiu, M. C. and Von-Mehlem, U. I. and Willey, C. E. and Betenbaugh, T. M. and Maynard, J. J. and Krein, J. A. and Conde, R. F. and Gray, W. T. and Hunt, J. W. and Mosher, L. E. and McCullough, M. G. and Panneton, P. E. and Staiger, J. P. and Rodberg, E. H.",
    editor="Russell, C. T. and Mewaldt, R. A.
and Von Rosenvinge, T. T.",
    title="Ace Spacecraft",
    bookTitle="The Advanced Composition Explorer Mission",
    year="1998",
publisher="Springer Netherlands",
    address="Dordrecht",
    pages="257--284",
    isbn="978-94-011-4762-0",
    doi="10.1007/978-94-011-4762-0_13",
}

@article{EPAM_Gold_Krimigis_Hawkins_Haggerty_Lohr_Fiore_Armstrong_Holland_Lanzerotti_1998, 
    title={Electron, Proton, and Alpha Monitor on the Advanced Composition Explorer spacecraft}, volume={86}, 
    ISSN={1572-9672}, 
    DOI={10.1023/A:1005088115759}, 
    number={1}, 
    journal={Space Science Reviews}, 
    author={Gold, R.E. and Krimigis, S.M. and Hawkins, S.E. and Haggerty, D.K. and Lohr, D.A. and Fiore, E. and Armstrong, T.P. and Holland, G. and Lanzerotti, L.J.}, year={1998}, 
    month=jul, 
    pages={541–562} 
}

@ARTICLE{MAG_Smith_1998,
       author = {{Smith}, C.~W. and {L'Heureux}, J. and {Ness}, N.~F. and {Acu{\~n}a}, M.~H. and {Burlaga}, L.~F. and {Scheifele}, J.},
        title = "{The ACE Magnetic Fields Experiment}",
      journal = {Space Science Reviews},
         year = 1998,
        month = jul,
       volume = {86},
        pages = {613-632},
          doi = {10.1023/A:1005092216668},
}

@article{SIS_Stone_1998, 
    title={The Solar Isotope Spectrometer for the Advanced Composition Explorer}, 
    volume={86}, 
    ISSN={1572-9672}, 
    DOI={10.1023/A:1005027929871}, 
    number={1}, 
    journal={Space Science Reviews}, 
    author={Stone, E.C. and Cohen, C.M.S. and Cook, W.R. and Cummings, A.C. and Gauld, B. and Kecman, B. and Leske, R.A. and Mewaldt, R.A. and Thayer, M.R. and Dougherty, B.L. and Grumm, R.L. and Milliken, B.D. and Radocinski, R.G. and Wiedenbeck, M.E. and Christian, E.R. and Shuman, S. and von Rosenvinge, T.T.}, 
    year={1998}, 
    month=jul, 
    pages={357–408} 
}

@article{SWEPAM_Comas_1998, 
    title={Solar Wind Electron Proton Alpha Monitor (SWEPAM) for the Advanced Composition Explorer}, 
    volume={86}, ISSN={1572-9672}, 
    DOI={10.1023/A:1005040232597}, 
    number={1}, 
    journal={Space Science Reviews}, 
    author={McComas, D.J. and Bame, S.J. and Barker, P. and Feldman, W.C. and Phillips, J.L. and Riley, P. and Griffee, J.W.}, year={1998}, 
    month=jul, 
    pages={563–612} 
}

@INPROCEEDINGS{DSCOVR,
  author={Burt, Joe and Smith, Bob},
  booktitle={2012 IEEE Aerospace Conference}, 
  title={Deep Space Climate Observatory: The DSCOVR mission}, 
  year={2012},
  volume={},
  number={},
  pages={1-13},
  doi={10.1109/AERO.2012.6187025}
}

@ARTICLE{al-haddad_2025_cme_repr,
       author = {{Al-Haddad}, Nada and {Lugaz}, No{\'e}},
        title = "{The Magnetic Field Structure of Coronal Mass Ejections: A More Realistic Representation}",
      journal = {\ssr},
         year = 2025,
        month = feb,
       volume = {221},
       number = {1},
          eid = {12},
        pages = {12},
          doi = {10.1007/s11214-025-01138-w},
       adsurl = {https://ui.adsabs.harvard.edu/abs/2025SSRv..221...12A}
}

@article{bernoux_2022_geomagforecast,
    author = {Bernoux, Guillerme and Brunet, Antoine and Buchlin, Eric and Janvier, Miho and Sicard, Angélica},
    title = {Forecasting the Geomagnetic Activity Several Days in Advance Using Neural Networks Driven by Solar EUV Imaging},
    journal = {Journal of Geophysical Research: Space Physics},
    volume = {127},
    number = {10},
    pages = {e2022JA030868},
    doi = {https://doi.org/10.1029/2022JA030868},
    year = {2022}
}

@article{Bilski_2023_calimera,
    title={CALIMERA: A New Early Time Series Classification Method},
    volume={60}, 
    ISSN={0306-4573}, 
    DOI={10.1016/j.ipm.2023.103465}, 
    number={5}, 
    journal={Information Processing \& Management}, 
    author={Bilski, Jakub Michal and Jastrzebska, Agnieszka}, 
    year={2023}, 
    month=sep, 
    pages={103465}
}

@article{Bouriat_2022, 
    title={Towards an AI-based understanding of the solar wind: A critical data analysis of ACE data}, 
    volume={9}, 
    ISSN={2296-987X}, 
    DOI={10.3389/fspas.2022.980759}, 
    journal={Frontiers in Astronomy and Space Sciences},
    author={Bouriat, S. and Vandame, P. and Barthélémy, M. and Chanussot, J.}, 
    year={2022}, 
    month=nov, 
    language={English} 
}

@ARTICLE{burlaga_1981,
       author = {{Burlaga}, L. and {Sittler}, E. and {Mariani}, F. and {Schwenn}, R.},
        title = "{Magnetic loop behind an interplanetary shock: Voyager, Helios, and IMP 8 observations}",
      journal = {Journal of Geophysics Research},
         year = 1981,
        month = aug,
       volume = {86},
       number = {A8},
        pages = {6673-6684},
          doi = {10.1029/JA086iA08p06673},
}

@article{camporeale_classification_2017,
	title = {Classification of {Solar} {Wind} {With} {Machine} {Learning}},
	volume = {122},
	copyright = {©2017. American Geophysical Union. All Rights Reserved.},
	issn = {2169-9402},
	doi = {10.1002/2017JA024383},
	language = {en},
	number = {11},
	journal = {Journal of Geophysical Research: Space Physics},
	author = {Camporeale, Enrico and Carè, Algo and Borovsky, Joseph E.},
	year = {2017},
	pages = {10,910--10,920},
}

@article{chen_ru-net_2022,
	title = {{RU}-net: {A} {Residual} {U}-net for {Automatic} {Interplanetary} {Coronal} {Mass} {Ejection} {Detection}},
	volume = {259},
	issn = {0067-0049, 1538-4365},
	shorttitle = {{RU}-net},
	doi = {10.3847/1538-4365/ac4587},
	number = {1},
	journal = {The Astrophysical Journal Supplement Series},
	author = {Chen, Jun and Deng, Hao and Li, Shuxin and Li, Weifu and Chen, Hong and Chen, Yanhong and Luo, Bingxian},
	month = mar,
	year = {2022},
	pages = {8},
}

@ARTICLE{Chi_2016,
       author = {{Chi}, Yutian and {Shen}, Chenglong and {Wang}, Yuming and {Xu}, Mengjiao and {Ye}, Pinzhong and {Wang}, Shui},
        title = "{Statistical Study of the Interplanetary Coronal Mass Ejections from 1995 to 2015}",
      journal = {Solar Physics},
         year = 2016,
        month = oct,
       volume = {291},
       number = {8},
        pages = {2419-2439},
          doi = {10.1007/s11207-016-0971-5},
archivePrefix = {arXiv},
       eprint = {1504.07849},
 primaryClass = {astro-ph.SR},
}

@article{Dachraoui_2015_earlyclass, 
    address={Cham}, title={Early Classification of Time Series as a Non Myopic Sequential Decision Making Problem}, 
    ISBN={978-3-319-23528-8}, 
    DOI={10.1007/978-3-319-23528-8_27}, 
    booktitle={Machine Learning and Knowledge Discovery in Databases},
    publisher={Springer International Publishing}, 
    author={Dachraoui, Asma and Bondu, Alexis and Cornuéjols, Antoine}, 
    editor={Appice, Annalisa and Rodrigues, Pedro Pereira and Santos Costa, Vítor and Soares, Carlos and Gama, João and Jorge, Alípio}, 
    year={2015}, 
    pages={433–447}
}

@ARTICLE{davies_2022_multispacecraft,
       author = {{Davies}, Emma E. and {Winslow}, R{\'e}ka M. and {Scolini}, Camilla and {Forsyth}, Robert J. and {M{\"o}stl}, Christian and {Lugaz}, No{\'e} and {Galvin}, Antoinette B.},
        title = "{Multi-spacecraft Observations of the Evolution of Interplanetary Coronal Mass Ejections between 0.3 and 2.2 au: Conjunctions with the Juno Spacecraft}",
      journal = {\apj},
         year = 2022,
        month = jul,
       volume = {933},
       number = {2},
          eid = {127},
        pages = {127},
          doi = {10.3847/1538-4357/ac731a},
archivePrefix = {arXiv},
       eprint = {2205.09472},
 primaryClass = {physics.space-ph},
       adsurl = {https://ui.adsabs.harvard.edu/abs/2022ApJ...933..127D}
}

@ARTICLE{davies_2025_realtimepred,
       author = {{Davies}, Emma E. and {Weiler}, Eva and {M{\"o}stl}, Christian and {Horbury}, Timothy S. and {O'Brien}, Helen and {Morris}, Jean and {Crabtree}, Alastair},
        title = "{Real-time prediction of geomagnetic storms using Solar Orbiter as a far upstream solar wind monitor}",
      journal = {arXiv e-prints},
         year = 2025,
        month = aug,
          eid = {arXiv:2508.13892},
        pages = {arXiv:2508.13892},
          doi = {10.48550/arXiv.2508.13892},
archivePrefix = {arXiv},
       eprint = {2508.13892},
 primaryClass = {physics.space-ph},
       adsurl = {https://ui.adsabs.harvard.edu/abs/2025arXiv250813892D}
}

@article{echer_2013,
    author = {Echer, E. and Tsurutani, B. T. and Gonzalez, W. D.},
    title = {Interplanetary origins of moderate ({-100} nT < {Dst} $\leq$ {-50} nT) geomagnetic storms during solar cycle 23 (1996--2008)},
    journal = {Journal of Geophysical Research: Space Physics},
    volume = {118},
    number = {1},
    pages = {385--392},
    year = {2013},
    doi = {10.1029/2012JA018086},
    url = {https://agupubs.onlinelibrary.wiley.com/doi/abs/10.1029/2012JA018086},
    eprint = {https://agupubs.onlinelibrary.wiley.com/doi/pdf/10.1029/2012JA018086}
}

@article{farooki_machine_2024,
	title = {A {Machine} {Learning} {Approach} to {Understanding} the {Physical} {Properties} of {Magnetic} {Flux} {Ropes} in the {Solar} {Wind} at 1 au},
	volume = {961},
	issn = {0004-637X},
	doi = {10.3847/1538-4357/ad0c52},
	language = {en},
	number = {1},
	journal = {The Astrophysical Journal},
	author = {Farooki, Hameedullah and Abduallah, Yasser and Noh, Sung Jun and Kim, Hyomin and Bizos, George and Shin, Youra and Wang, Jason T. L. and Wang, Haimin},
	month = jan,
	year = {2024},
	pages = {81},
}

@ARTICLE{good2018correlation,
       author = {{Good}, S.~W. and {Forsyth}, R.~J. and {Eastwood}, J.~P. and {M{\"o}stl}, C.},
        title = "{Correlation of ICME Magnetic Fields at Radially Aligned Spacecraft}",
      journal = {Solar Physics},
         year = 2018,
        month = mar,
       volume = {293},
       number = {3},
          eid = {52},
        pages = {52},
          doi = {10.1007/s11207-018-1264-y},
archivePrefix = {arXiv},
       eprint = {1802.04004},
 primaryClass = {physics.space-ph},
}

@ARTICLE{gosling_1973,
       author = {{Gosling}, J.~T. and {Pizzo}, V. and {Bame}, S.~J.},
        title = "{Anomalously low proton temperatures in the solar wind following interplanetary shock waves{\textemdash}evidence for magnetic bottles?}",
      journal = {Journal of Geophysics Research},
         year = 1973,
        month = jan,
       volume = {78},
       number = {13},
        pages = {2001},
          doi = {10.1029/JA078i013p02001},
}

@article{Hu_2018,
    doi = {10.3847/1538-4365/aae57d},
    url = {https://dx.doi.org/10.3847/1538-4365/aae57d},
    year = {2018},
    month = {nov},
    publisher = {The American Astronomical Society},
    volume = {239},
    number = {1},
    pages = {12},
    author = {Hu, Qiang and Zheng, Jinlei and Chen, Yu and le Roux, Jakobus and Zhao, Lulu},
    title = {Automated Detection of Small-scale Magnetic Flux Ropes in the Solar Wind: First Results from the Wind Spacecraft Measurements},
    journal = {The Astrophysical Journal Supplement Series},
}

@INPROCEEDINGS{jadon_2020,
    author={Jadon, Shruti},
    booktitle={2020 IEEE Conference on Computational Intelligence in Bioinformatics and Computational Biology (CIBCB)}, 
    title={A survey of loss functions for semantic segmentation}, 
    year={2020},
    volume={},
    number={},
    pages={1-7},
    doi={10.1109/CIBCB48159.2020.9277638}
}

@ARTICLE{janvier_2019_generic,
       author = {{Janvier}, Miho and {Winslow}, Reka M. and {Good}, Simon and {Bonhomme}, Elise and {D{\'e}moulin}, Pascal and {Dasso}, Sergio and {M{\"o}stl}, Christian and {Lugaz}, No{\'e} and {Amerstorfer}, Tanja and {Soubri{\'e}}, Elie and {Boakes}, Peter D.},
        title = "{Generic Magnetic Field Intensity Profiles of Interplanetary Coronal Mass Ejections at Mercury, Venus, and Earth From Superposed Epoch Analyses}",
      journal = {Journal of Geophysical Research (Space Physics)},
         year = 2019,
        month = feb,
       volume = {124},
       number = {2},
        pages = {812-836},
          doi = {10.1029/2018JA025949},
archivePrefix = {arXiv},
       eprint = {1901.09921},
 primaryClass = {physics.space-ph},
       adsurl = {https://ui.adsabs.harvard.edu/abs/2019JGRA..124..812J}
}

@ARTICLE{jian_2006,
       author = {{Jian}, L. and {Russell}, C.~T. and {Luhmann}, J.~G. and {Skoug}, R.~M.},
        title = "{Properties of Interplanetary Coronal Mass Ejections at One AU During 1995   2004}",
      journal = {Solar Physics},
         year = 2006,
        month = dec,
       volume = {239},
       number = {1-2},
        pages = {393-436},
          doi = {10.1007/s11207-006-0133-2},
}

@ARTICLE{kilpua2009,
   author = {{Kilpua}, E.~K.~J. and {Liewer}, P.~C. and {Farrugia}, C. and 
	{Luhmann}, J.~G. and {M{\"o}stl}, C. and {Li}, Y. and {Liu}, Y. and 
	{Lynch}, B.~J. and {Russell}, C.~T. and {Vourlidas}, A. and 
	{Acuna}, M.~H. and {Galvin}, A.~B. and {Larson}, D. and {Sauvaud}, J.~A.
	},
    title = "{Multispacecraft Observations of Magnetic Clouds and Their Solar Origins between 19 and 23 May 2007}",
  journal = {Solar Physics},
     year = 2009,
    month = feb,
   volume = 254,
    pages = {325-344},
      doi = {10.1007/s11207-008-9300-y},
}

@ARTICLE{kilpua_2017,
       author = {{Kilpua}, Emilia and {Koskinen}, Hannu E.~J. and {Pulkkinen}, Tuija I.},
        title = "{Coronal mass ejections and their sheath regions in interplanetary space}",
      journal = {Living Reviews in Solar Physics},
         year = 2017,
        month = dec,
       volume = {14},
       number = {1},
          eid = {5},
        pages = {5},
          doi = {10.1007/s41116-017-0009-6},
}

@article{King_Papitashvili_2005, 
    title={Solar wind spatial scales in and comparisons of hourly Wind and ACE plasma and magnetic field data}, 
    volume={110}, 
    rights={Copyright 2005 by the American Geophysical Union.}, 
    ISSN={2156-2202}, 
    DOI={10.1029/2004JA010649}, 
    number={A2}, 
    journal={Journal of Geophysical Research: Space Physics}, 
    author={King, J. H. and Papitashvili, N. E.}, 
    year={2005}, 
    language={en} 
}

@article{kingma2014_adam,
    author = {Kingma, Diederik and Ba, Jimmy},
    year = {2014},
    month = {12},
    pages = {},
    title = {Adam: A Method for Stochastic Optimization},
    journal = {International Conference on Learning Representations}
}

@ARTICLE{klein_1982,
       author = {{Klein}, L.~W. and {Burlaga}, L.~F.},
        title = "{Interplanetary magnetic clouds at 1 AU}",
      journal = {Journal of Geophysics Research},
         year = 1982,
        month = feb,
       volume = {87},
       number = {A2},
        pages = {613-624},
          doi = {10.1029/JA087iA02p00613},
}

@Article{lepping_2005_automaticdetection,
    AUTHOR = {Lepping, R. P. and Wu, C.-C. and Berdichevsky, D. B.},
    TITLE = {Automatic identification of magnetic clouds and cloud-like regions at 1 AU: occurrence rate and other properties},
    JOURNAL = {Annales Geophysicae},
    VOLUME = {23},
    YEAR = {2005},
    NUMBER = {7},
    PAGES = {2687--2704},
    DOI = {10.5194/angeo-23-2687-2005}
}

@ARTICLE{lepping_2006,
       author = {{Lepping}, R.~P. and {Berdichevsky}, D.~B. and {Wu}, C. -C. and {Szabo}, A. and {Narock}, T. and {Mariani}, F. and {Lazarus}, A.~J. and {Quivers}, A.~J.},
        title = "{A summary of WIND magnetic clouds for years 1995-2003: model-fitted parameters, associated errors and classifications}",
      journal = {Annales Geophysicae},
         year = 2006,
        month = mar,
       volume = {24},
       number = {1},
        pages = {215-245},
          doi = {10.5194/angeo-24-215-2006},
}

@ARTICLE{laker_2024,
       author = {{Laker}, R. and {Horbury}, T.~S. and {O'Brien}, H. and {Fauchon-Jones}, E.~J. and {Angelini}, V. and {Fargette}, N. and {Amerstorfer}, T. and {Bauer}, M. and {M{\"o}stl}, C. and {Davies}, E.~E. and {Davies}, J.~A. and {Harrison}, R. and {Barnes}, D. and {Dumbovi{\'c}}, M.},
        title = "{Using Solar Orbiter as an Upstream Solar Wind Monitor for Real Time Space Weather Predictions}",
      journal = {Space Weather},
         year = 2024,
        month = feb,
       volume = {22},
       number = {2},
          eid = {e2023SW003628},
        pages = {e2023SW003628},
          doi = {10.1029/2023SW003628},
archivePrefix = {arXiv},
       eprint = {2307.01083},
 primaryClass = {physics.space-ph},
       adsurl = {https://ui.adsabs.harvard.edu/abs/2024SpWea..2203628L}
}

@ARTICLE{li_categorization_2020,
       author = {{Li}, Hui and {Wang}, Chi and {Tu}, Cui and {Xu}, Fei},
        title = "{Machine Learning Approach for Solar Wind Categorization}",
      journal = {Earth and Space Science},
         year = 2020,
        month = may,
       volume = {7},
       number = {5},
          eid = {e00997},
        pages = {e00997},
          doi = {10.1029/2019EA000997},
archivePrefix = {arXiv},
       eprint = {1811.02323},
 primaryClass = {physics.space-ph},
       adsurl = {https://ui.adsabs.harvard.edu/abs/2020E&SS....700997L}
}

@article{lotoaniu_validation_2022,
	title = {Validation of the {DSCOVR} {Spacecraft} {Mission} {Space} {Weather} {Solar} {Wind} {Products}},
	volume = {20},
	copyright = {© 2022. The Authors.},
	issn = {1542-7390},
	doi = {10.1029/2022SW003085},
	language = {en},
	number = {10},
	journal = {Space Weather},
	author = {Loto'aniu, Paul T. M. and Romich, K. and Rowland, W. and Codrescu, S. and Biesecker, D. and Johnson, J. and Singer, H. J. and Szabo, A. and Stevens, M.},
	year = {2022},
	pages = {e2022SW003085},
}

@article{Lugaz_2018, 
    title={On the Spatial Coherence of Magnetic Ejecta: Measurements of Coronal Mass Ejections by Multiple Spacecraft Longitudinally Separated by 0.01 au}, 
    volume={864}, 
    ISSN={2041-8205}, 
    DOI={10.3847/2041-8213/aad9f4}, 
    number={1}, 
    journal={The Astrophysical Journal Letters},
    publisher={The American Astronomical Society}, 
    author={Lugaz, Noé and Farrugia, Charles J. and Winslow, Reka M. and Al-Haddad, Nada and Galvin, Antoinette B. and Nieves-Chinchilla, Teresa and Lee, Christina O. and Janvier, Miho}, 
    year={2018}, 
    month=aug, 
    pages={L7}, 
    language={en} 
}

@ARTICLE{lugaz_2024,
       author = {{Lugaz}, No{\'e} and {Lee}, Christina O. and {Al-Haddad}, Nada and {Lillis}, Robert J. and {Jian}, Lan K. and {Curtis}, David W. and {Galvin}, Antoinette B. and {Whittlesey}, Phyllis L. and {Rahmati}, Ali and {Zesta}, Eftyhia and {Moldwin}, Mark and {Summerlin}, Errol J. and {Larson}, Davin E. and {Courtade}, Sasha and {French}, Richard and {Hunter}, Richard and {Covitti}, Federico and {Cosgrove}, Daniel and {Prall}, J.~D. and {Allen}, Robert C. and {Zhuang}, Bin and {Winslow}, R{\'e}ka M. and {Scolini}, Camilla and {Lynch}, Benjamin J. and {Filwett}, Rachael J. and {Palmerio}, Erika and {Farrugia}, Charles J. and {Smith}, Charles W. and {M{\"o}stl}, Christian and {Weiler}, Eva and {Janvier}, Miho and {Regnault}, Florian and {Livi}, Roberto and {Nieves-Chinchilla}, Teresa},
        title = "{The Need for Near-Earth Multi-Spacecraft Heliospheric Measurements and an Explorer Mission to Investigate Interplanetary Structures and Transients in the Near-Earth Heliosphere}",
      journal = {Space Science Reviews},
         year = 2024,
        month = oct,
       volume = {220},
       number = {7},
          eid = {73},
        pages = {73},
          doi = {10.1007/s11214-024-01108-8},
       adsurl = {https://ui.adsabs.harvard.edu/abs/2024SSRv..220...73L}
}

@ARTICLE{moestl2015elevo,
       author = {{M{\"o}stl}, Christian and {Rollett}, Tanja and {Frahm}, Rudy A. and {Liu}, Ying D. and {Long}, David M. and {Colaninno}, Robin C. and {Reiss}, Martin A. and {Temmer}, Manuela and {Farrugia}, Charles J. and {Posner}, Arik and {Dumbovi{\'c}}, Mateja and {Janvier}, Miho and {D{\'e}moulin}, Pascal and {Boakes}, Peter and {Devos}, Andy and {Kraaikamp}, Emil and {Mays}, Mona L. and {Vr{\v{s}}nak}, Bojan},
        title = "{Strong coronal channelling and interplanetary evolution of a solar storm up to Earth and Mars}",
      journal = {Nature Communications},
         year = 2015,
        month = may,
       volume = {6},
          eid = {7135},
        pages = {7135},
          doi = {10.1038/ncomms8135},
archivePrefix = {arXiv},
       eprint = {1506.02842},
 primaryClass = {astro-ph.SR},
       adsurl = {https://ui.adsabs.harvard.edu/abs/2015NatCo...6.7135M}
}

@ARTICLE{moestl_2017,
       author = {{M{\"o}stl}, C. and {Isavnin}, A. and {Boakes}, P.~D. and {Kilpua}, E.~K.~J. and {Davies}, J.~A. and {Harrison}, R.~A. and {Barnes}, D. and {Krupar}, V. and {Eastwood}, J.~P. and {Good}, S.~W. and {Forsyth}, R.~J. and {Bothmer}, V. and {Reiss}, M.~A. and {Amerstorfer}, T. and {Winslow}, R.~M. and {Anderson}, B.~J. and {Philpott}, L.~C. and {Rodriguez}, L. and {Rouillard}, A.~P. and {Gallagher}, P. and {Nieves-Chinchilla}, T. and {Zhang}, T.~L.},
        title = "{Modeling observations of solar coronal mass ejections with heliospheric imagers verified with the Heliophysics System Observatory}",
      journal = {Space Weather},
         year = 2017,
        month = jul,
       volume = {15},
       number = {7},
        pages = {955-970},
          doi = {10.1002/2017SW001614},
archivePrefix = {arXiv},
       eprint = {1703.00705},
 primaryClass = {astro-ph.SR},
}

@article{moestl2020_icmerate,
    doi = {10.3847/1538-4357/abb9a1},
    url = {https://dx.doi.org/10.3847/1538-4357/abb9a1},
    year = {2020},
    month = {nov},
    publisher = {The American Astronomical Society},
    volume = {903},
    number = {2},
    pages = {92},
    author = {Möstl, Christian and Weiss, Andreas J. and Bailey, Rachel L. and Reiss, Martin A. and Amerstorfer, Tanja and Hinterreiter, Jürgen and Bauer, Maike and McIntosh, Scott W. and Lugaz, Noé and Stansby, David},
    title = {Prediction of the In Situ Coronal Mass Ejection Rate for Solar Cycle 25: Implications for Parker Solar Probe In Situ Observations},
    journal = {The Astrophysical Journal},
}

@DATA{moestl2025_catalog,
author = "Christian Möstl and Emma Davies and Eva Weiler",
title = "{HELIO4CAST Interplanetary Coronal Mass Ejection Catalog v2.3}",
year = "2025",
month = "7",
url = "https://figshare.com/articles/dataset/HELCATS_Interplanetary_Coronal_Mass_Ejection_Catalog_v2_0/6356420",
doi = "https://doi.org/10.6084/m9.figshare.6356420.v23"
}

@article{narock_classifying_2024,
	title = {Classifying {Different} {Types} of {Solar}-{Wind} {Plasma} with {Uncertainty} {Estimations} {Using} {Machine} {Learning}},
	volume = {299},
	issn = {1573-093X},
	doi = {10.1007/s11207-024-02379-8},
	language = {en},
	number = {9},
	journal = {Solar Physics},
	author = {Narock, Tom and Pal, Sanchita and Arsham, Aryana and Narock, Ayris and Nieves-Chinchilla, Teresa},
	month = sep,
	year = {2024},
	pages = {131},
}

@article{nguyen_automatic_2019,
	title = {Automatic {Detection} of {Interplanetary} {Coronal} {Mass} {Ejections} from {In} {Situ} {Data}: {A} {Deep} {Learning} {Approach}},
	volume = {874},
	issn = {0004-637X, 1538-4357},
	shorttitle = {Automatic {Detection} of {Interplanetary} {Coronal} {Mass} {Ejections} from {In} {Situ} {Data}},
	doi = {10.3847/1538-4357/ab0d24},
	number = {2},
	journal = {The Astrophysical Journal},
	author = {Nguyen, Gautier and Aunai, Nicolas and Fontaine, Dominique and Pennec, Erwan Le and Bossche, Joris Van Den and Jeandet, Alexis and Bakkali, Brice and Vignoli, Louis and Blancard, Bruno Regaldo-Saint},
	month = apr,
	year = {2019},
	pages = {145},
}

@article{nguyen_multiclass_2025,
	title = {Simultaneous multi-class detection of interplanetary space weather events},
	author = {Nguyen, Gautier and Bernoux, Guillerme and Ferlin, A.},
	year = {2025},
        journal = {Journal of Space Weather and Space Climate},
        month = apr,
        doi = {10.1051/swsc/2025016},

}

@ARTICLE{nieves_chinchilla_2018,
       author = {{Nieves-Chinchilla}, T. and {Vourlidas}, A. and {Raymond}, J.~C. and {Linton}, M.~G. and {Al-haddad}, N. and {Savani}, N.~P. and {Szabo}, A. and {Hidalgo}, M.~A.},
        title = "{Understanding the Internal Magnetic Field Configurations of ICMEs Using More than 20 Years of Wind Observations}",
      journal = {Solar Physics},
         year = 2018,
        month = feb,
       volume = {293},
       number = {2},
          eid = {25},
        pages = {25},
          doi = {10.1007/s11207-018-1247-z},
}

@article{ojedagonzalez2017,
    author = {Ojeda Gonzalez, Arian and Mendes, Odim and Calzadilla, Alexander and Domingues, Margarete and Prestes, Alan and Klausner, Virginia},
    year = {2017},
    month = {03},
    pages = {156},
    title = {An Alternative Method for Identifying Interplanetary Magnetic Cloud Regions},
    volume = {837},
    journal = {The Astrophysical Journal},
    doi = {10.3847/1538-4357/aa6034}
}

@article{pal_automatic_2024,
	title = {Automatic {Detection} of {Large}-scale {Flux} {Ropes} and {Their} {Geoeffectiveness} with a {Machine}-learning {Approach}},
	volume = {972},
	issn = {0004-637X},
	doi = {10.3847/1538-4357/ad54c3},
	language = {en},
	number = {1},
	journal = {The Astrophysical Journal},
	author = {Pal, Sanchita and Santos, Luiz F. G. dos and Weiss, Andreas J. and Narock, Thomas and Narock, Ayris and Nieves-Chinchilla, Teresa and Jian, Lan K. and Good, Simon W.},
	month = aug,
	year = {2024},
	pages = {94},
}

@INPROCEEDINGS{redmon_2016_yolo,
  author={Redmon, Joseph and Divvala, Santosh and Girshick, Ross and Farhadi, Ali},
  booktitle={2016 IEEE Conference on Computer Vision and Pattern Recognition (CVPR)}, 
  title={You Only Look Once: Unified, Real-Time Object Detection}, 
  year={2016},
  volume={},
  number={},
  pages={779-788},
  doi={10.1109/CVPR.2016.91}
}

@ARTICLE{richardson_2010,
       author = {{Richardson}, I.~G. and {Cane}, H.~V.},
        title = "{Near-Earth Interplanetary Coronal Mass Ejections During Solar Cycle 23 (1996 - 2009): Catalog and Summary of Properties}",
      journal = {Solar Physics},
         year = 2010,
        month = jun,
       volume = {264},
       number = {1},
        pages = {189-237},
          doi = {10.1007/s11207-010-9568-6},
}

@article{richardson_2012,
	author = {{Richardson, Ian G.} and {Cane, Hilary V.}},
	title = {Solar wind drivers of geomagnetic storms during more than four solar cycles},
	DOI= "10.1051/swsc/2012001",
	url= "https://doi.org/10.1051/swsc/2012001",
	journal = {J. Space Weather Space Clim.},
	year = 2012,
	volume = 2,
	pages = "A01",
}

@ARTICLE{richardson_2014_catdifferences,
       author = {{Richardson}, I.~G.},
        title = "{Identification of Interplanetary Coronal Mass Ejections at Ulysses Using Multiple Solar Wind Signatures}",
      journal = {\solphys},
         year = 2014,
        month = oct,
       volume = {289},
       number = {10},
        pages = {3843-3894},
          doi = {10.1007/s11207-014-0540-8},
       adsurl = {https://ui.adsabs.harvard.edu/abs/2014SoPh..289.3843R}
}

@ARTICLE{riley_2023,
       author = {{Riley}, Pete and {Reiss}, M.~A. and {M{\"o}stl}, C.},
        title = "{Which Upstream Solar Wind Conditions Matter Most in Predicting B$_{z}$ Within Coronal Mass Ejections}",
      journal = {Space Weather},
         year = 2023,
        month = apr,
       volume = {21},
       number = {4},
          eid = {e2022SW003327},
        pages = {e2022SW003327},
          doi = {10.1029/2022SW003327},
archivePrefix = {arXiv},
       eprint = {2303.17682},
 primaryClass = {physics.space-ph},
}

@article{rudisser_automatic_2022,
	title = {Automatic {Detection} of {Interplanetary} {Coronal} {Mass} {Ejections} in {Solar} {Wind} {In} {Situ} {Data}},
	volume = {20},
	issn = {1542-7390, 1542-7390},
	doi = {10.1029/2022SW003149},
	language = {en},
	number = {10},
	journal = {Space Weather},
	author = {Rüdisser, H. T. and Windisch, A. and Amerstorfer, U. V. and Möstl, C. and Amerstorfer, T. and Bailey, R. L. and Reiss, M. A.},
	month = oct,
	year = {2022},
	pages = {e2022SW003149},
}

@article{ruedisser_2024,
doi = {10.3847/1538-4357/ad660a},
url = {https://dx.doi.org/10.3847/1538-4357/ad660a},
year = {2024},
month = {sep},
publisher = {The American Astronomical Society},
volume = {973},
number = {2},
pages = {150},
author = {Hannah T. Rüdisser and Andreas J. Weiss and Justin Le Louëdec and Ute V. Amerstorfer and Christian Möstl and Emma E. Davies and Helmut Lammer},
title = {Understanding the Effects of Spacecraft Trajectories through Solar Coronal Mass Ejection Flux Ropes Using 3DCOREweb},
journal = {The Astrophysical Journal},
}

@software{ruedisser2025_data,
    author = "Hannah T. Rüdisser and Justin Le Louëdec and Gautier Nguyen and Emma E. Davies and Christian Möstl",
    title = "{ARCANE - Early Detection of Interplanetary Coronal Mass Ejections}",
    year = "2025",
    month = "1",
    url = "https://doi.org/10.6084/m9.figshare.28309295.v6",
    doi = "10.6084/m9.figshare.28309295.v6"
}

@article{salman_2018,
author = {Salman, T. M. and Lugaz, N. and Farrugia, C. J. and Winslow, R. M. and Galvin, A. B. and Schwadron, N. A.},
title = {Forecasting Periods of Strong Southward Magnetic Field Following Interplanetary Shocks},
journal = {Space Weather},
volume = {16},
number = {12},
pages = {2004-2021},
doi = {https://doi.org/10.1029/2018SW002056},
year = {2018}
}

@ARTICLE{salman2020radial,
       author = {{Salman}, T.~M. and {Winslow}, R.~M. and {Lugaz}, N.},
        title = "{Radial Evolution of Coronal Mass Ejections Between MESSENGER, Venus Express, STEREO, and L1: Catalog and Analysis}",
      journal = {Journal of Geophysics Research (Space Physics)},
         year = 2020,
        month = jan,
       volume = {125},
       number = {1},
          eid = {e27084},
        pages = {e27084},
          doi = {10.1029/2019JA027084},
archivePrefix = {arXiv},
       eprint = {1912.11731},
 primaryClass = {physics.space-ph},
}

@ARTICLE{smith2022,
       author = {{Smith}, A.~W. and {Forsyth}, C. and {Rae}, I.~J. and {Garton}, T.~M. and {Jackman}, C.~M. and {Bakrania}, M. and {Shore}, R.~M. and {Richardson}, G.~S. and {Beggan}, C.~D. and {Heyns}, M.~J. and {Eastwood}, J.~P. and {Thomson}, A.~W.~P. and {Johnson}, J.~M.},
        title = "{On the Considerations of Using Near Real Time Data for Space Weather Hazard Forecasting}",
      journal = {Space Weather},
         year = 2022,
        month = jul,
       volume = {20},
       number = {7},
          eid = {e2022SW003098},
        pages = {e2022SW003098},
          doi = {10.1029/2022SW003098},
}

@article{turner_solar_2023,
	title = {Solar {Wind} {Data} {Assimilation} in an {Operational} {Context}: {Use} of {Near}-{Real}-{Time} {Data} and the {Forecast} {Value} of an {L5} {Monitor}},
	volume = {21},
	copyright = {© 2023. The Authors.},
	issn = {1542-7390},
	shorttitle = {Solar {Wind} {Data} {Assimilation} in an {Operational} {Context}},
	doi = {10.1029/2023SW003457},
	language = {en},
	number = {5},
	journal = {Space Weather},
	author = {Turner, Harriet and Lang, Matthew and Owens, Mathew and Smith, Andy and Riley, Pete and Marsh, Mike and Gonzi, Siegfried},
	year = {2023},
	pages = {e2023SW003457},
}

@ARTICLE{wang_2025,
       author = {{Wang}, Jiye and {Liu}, Xuan and {Dai}, Fanzhuo and {Zheng}, Rui and {Han}, Yuanlin and {Wang}, Yang and {Liu}, Andi and {Wei}, Xinhua and {Zhang}, Lingqian and {Li}, Hui and {Wang}, Chi and {Wang}, Tieyan and {Burch}, James L. and {Baumjohann}, Wolfgang},
        title = "{Automated Plasma Region Classification and Boundary Layer Identification Using Machine Learning}",
      journal = {Remote Sensing},
         year = 2025,
        month = apr,
       volume = {17},
       number = {9},
          eid = {1565},
        pages = {1565},
          doi = {10.3390/rs17091565},
       adsurl = {https://ui.adsabs.harvard.edu/abs/2025RemS...17.1565W}
}

@ARTICLE{weiler_2025,
       author = {{Weiler}, Eva and {M{\"o}stl}, Christian and {Davies}, Emma E. and {Veronig}, Astrid and {Amerstorfer}, Ute V. and {Amerstorfer}, Tanja and {Le Lou{\"e}dec}, Justin and {Bauer}, Maike and {Lugaz}, No{\'e} and {Haberle}, Veronika and {R{\"u}disser}, Hannah T. and {Majumdar}, Satabdwa and {Reiss}, Martin},
        title = "{First observations of a geomagnetic superstorm with a sub-L1 monitor}",
      journal = {arXiv e-prints},
         year = 2024,
        month = nov,
          eid = {arXiv:2411.12490},
        pages = {arXiv:2411.12490},
          doi = {10.48550/arXiv.2411.12490},
archivePrefix = {arXiv},
       eprint = {2411.12490},
 primaryClass = {physics.space-ph},
       adsurl = {https://ui.adsabs.harvard.edu/abs/2024arXiv241112490W}
}

@article{Zafar_2021_earlyclass, 
    title={Early Classification of Time Series: Cost-based Multiclass Algorithms}, 
    DOI={10.1109/DSAA53316.2021.9564134}, 
    booktitle={2021 IEEE 8th International Conference on Data Science and Advanced Analytics (DSAA)}, 
    author={Zafar, Paul-Emile and Achenchabe, Youssef and Bondu, Alexis and Cornuéjols, Antoine and Lemaire, Vincent}, 
    year={2021}, 
    month=oct, 
    pages={1–10}
}

@article{zwickl_noaa_1998,
	title = {The {NOAA} {Real}-{Time} {Solar}-{Wind} ({RTSW}) {System} using {ACE} {Data}},
	volume = {86},
	issn = {1572-9672},
	doi = {10.1023/A:1005044300738},
	number = {1},
	journal = {Space Science Reviews},
	author = {Zwickl, R.D. and Doggett, K.A. and Sahm, S. and Barrett, W.P. and Grubb, R.N. and Detman, T.R. and Raben, V.J. and Smith, C.W. and Riley, P. and Gold, R.E. and Mewaldt, R.A. and Maruyama, T.},
	month = jul,
	year = {1998},
	pages = {633--648},
}

@ARTICLE{zurbuchen_2006,
       author = {{Zurbuchen}, Thomas H. and {Richardson}, Ian G.},
        title = "{In-Situ Solar Wind and Magnetic Field Signatures of Interplanetary Coronal Mass Ejections}",
      journal = {Space Science Reviews},
         year = 2006,
        month = mar,
       volume = {123},
       number = {1-3},
        pages = {31-43},
          doi = {10.1007/s11214-006-9010-4},
}

\end{document}